\documentclass[a4paper,fleqn,usenatbib]{mnras}

\usepackage{mathptmx}
\usepackage[T1]{fontenc}
\usepackage{ae,aecompl}

\usepackage{graphicx}	% Including figure files
\usepackage{amsmath}	% Advanced maths commands
\usepackage{amssymb}	% Extra maths symbols
\usepackage[export]{adjustbox}

\title[Interacting binary stars.]{AGB winds in interacting binary stars}

\author[L. C. Berm\'{u}dez et al.]{
Luis C. Berm\'{u}dez-Bustamante,\thanks{E-mail: luisb@astro.unam.mx}
G. Garc\'{i}a-Segura,
W. Steffen
and L.~Sabin
\\
Instituto de Astronom\'{i}a, Universidad Nacional Aut\'{o}noma de M\'{e}xico, km 107 Carr. Tijuana-Ensenada, Ensenada, B. C., 22860, M\'{e}xico\\
}

\date{Accepted XXX. Received YYY; in original form ZZZ}

\pubyear{2015}

\begin{document}
\label{firstpage}
\pagerange{\pageref{firstpage}--\pageref{lastpage}}
\maketitle

\begin{abstract}
We perform numerical simulations to investigate the stellar wind from interacting binary stars. Our aim is to find analytical formulae describing the outflow structure. In each binary system the more massive star is in the asymptotic giant branch and its wind is driven by a combination of pulsations in the stellar surface layers and radiation pressure on dust, while the less massive star is in the main sequence. Time averages of density and outflow velocity of the stellar wind are calculated and plotted as profiles against distance from the centre of mass and colatitude angle. We find that mass is lost mainly through the outer Lagrangian point $L_2$. The resultant outflow develops into a spiral at low distances from the binary. The outflowing spiral is quickly smoothed out by shocks and becomes an excretion disk at larger distances. This leads to the formation of an outflow structure with an equatorial density excess, which is greater in binaries with smaller orbital separation. The pole-to-equator density  ratio reaches a maximum value of $\sim10^5$ at Roche-Lobe Overflow state. We also find that the gas stream leaving $L_2$ does not form a circumbinary ring for stellar mass ratios above $0.78$, when radiation pressure on dust is taken into account. Analytical formulae are obtained by curve fitting the 2-dimensional, azimuthally averaged density and outflow velocity profiles. The formulae can be used in future studies to setup the initial outflow structure in hydrodynamic simulations of common-envelope evolution and formation of planetary nebulae.

\end{abstract}


\section{Introduction}

Single, low- and intermediate-mass stars lose their envelopes through slow, dense winds at the asymptotic giant branch (AGB) phase. At later stages, the fast and tenuous winds originating from the remnant stars interact with the inner layers of the expanding envelopes. This leads to the formation of planetary nebulae (PNe) once the remnant stars evolve towards higher effective temperatures and ionize the circumstellar gas \citep{1978ApJ...219L.125K,2000oepn.book.....K}. For the case of binary stars with the same mass range, the envelope ejection could be more violent, either at the red giant branch phase or at the AGB phase, if the system goes through a common envelope (CE). 
In both cases, the envelope ejection determines the distribution of gas in the circumstellar medium.

A fascinating, still unsolved problem of PNe is their morphological diversity, specially the bipolar class, a likely outcome of mass-loss geometry at the tip of the AGB \citep{1987AJ.....94..671B,1995A&A...293..871C,1996iacm.book.....M,2007AJ....134.2200S}.
Some numerical computations in the context of the interacting stellar winds model \citep{2002ASPC..260..245K} have reproduced a wide diversity of nebular shapes, but they assume arbitrary functions for the density and outflow velocity distribution of the slow AGB wind \citep{1991A&A...252..718M, 1992A&A...253..224I}.
A bipolar nebula can be formed if the density of the slow AGB wind increases towards the equatorial plane \citep{1985MNRAS.212..837K,1987AJ.....94..671B,1989AJ.....97..462I,1989ApJ...339..268S,1996ApJ...457..773D,1999ApJ...517..767G,2012MNRAS.424.2055H}. 
What causes this pole-to-equator density ratio in the AGB wind is a matter of discussion, but a stellar companion is the most promising shaping mechanism \citep{2014ApJ...783...74G}. 
In fact, $\sim40\%$  of the symbiotic stars have bipolar nebulae \citep{2004ASPC..313..497S}, and morpho-kinematical models of some PNe show that their symmetry axes are perpendicular to the orbital plane of their binary cores \citep[e.g.][]{2012MNRAS.420.2271J}. The extremely low probability for this perpendicularity to be a random event \citep{2016ApJ...832..125H}, shows a physical link between binary stars and nebular morphology.

Binary nuclei of PNe with orbital separations smaller than the typical radius of their progenitor AGB component suggest a loss of mass and angular momentum from the stellar systems in previous evolutionary stages.  
The basic formation scenario for close binaries  
begins when the primary star fills its Roche lobe during the ascent towards the AGB phase and transfers mass at dynamical timescales to a main-sequence companion (secondary star).
The latter cannot accrete or expel rapidly enough the donated material, from which a CE is formed \citep{2013A&ARv..21...59I}.
Gravitational drag transports orbital energy and angular momentum from the stellar system to the CE and reduces the orbital separation. 
The envelope is ejected leaving a tight binary star if sufficient energy is transferred, otherwise a stellar merger will occur.  
The formation of aspherical PNe by CE evolution is not fully understood \citep{2018ApJ...860...19G}, given that the details of the ejection mechanisms or their efficiency are unknown \citep{2017MNRAS.472.4361S}, and observations of this brief stellar phase \citep{2019MNRAS.486.1070C} are difficult. 

With the current computing power, hydrodynamic simulations of the entire CE phase are not possible, owing to its three-dimensional geometry and the wide range of length and time scales of the physical processes involved \citep{2019MNRAS.486.5809P}. In consequence numerical studies placed an emphasis on the dynamical spiral-in phase \citep{2012ApJ...746...74R,2012ApJ...744...52P}. 
However, initial conditions in computations of CE evolution may affect the outcome, specially when the orbital separation shrinks on dynamical timescales, not giving enough time for dissipation of the density structures formed immediately before. Therefore, a constant-density background gas in pressure equilibrium with the surface of the primary star at the onset of simulations \citep{2012ApJ...746...74R, 2016MNRAS.455.3511S}, or placing the companion star on the surface of the red giant \citep{2016ApJ...816L...9O} may not be realistic situations. 
In order to improve computations of CE evolution, specially the interaction of the CE ejecta with the previous circumbinary medium, the structure of the gas around the binary system has to be realistically calculated. Thus, the interaction of the AGB wind with the secondary star is very important in this context.

The effects of a stellar companion in the hydrodynamics of the AGB wind were previously investigated by \citet{1999ApJ...523..357M, 2002A&A...385..205G, 2007ASPC..372..397M,  2017MNRAS.468.4465C}, considering only detached binaries.
For the above cases, the AGB wind can be simple, well modelled as a supersonic flow, with a constant expansion velocity of the order of  $15~$km\,s$^{-1}$. 
However, for the case of semi-detached binary stars, the computation of the AGB wind has to be solved from the stellar surface, since the orbit of the binary lies inside the region where the AGB wind is accelerated. 
Thus, much more physics needs to be included in the modelling, and a more complete knowledge of the AGB wind has to be taken into account. In other words, the form in which the gas is ejected from the stellar surface and is accelerated up to the asymptotic expansion velocity has to be explicitly computed. 

Observations of molecular emission lines from AGB stars show that many of them undergo high mass-loss rates through slow, dense winds \citep{2010A&A...523A..18D,2018A&A...613L...7W}.
Unfortunately, for late-type giant stars, a complete explanation of the mass-loss mechanism is still missing. 
Consequently, estimations of the mass-loss rates must come from observations.
However, it is well known that AGB winds have asymptotic outflow speeds lower than the stellar surface escape velocity \citep{2009A&A...499..515R}, in comparison with less evolved Red Giant Branch stars \citep[e.g.][]{2018ApJ...869..157C}. This indicates that work is done by the mass-loss mechanisms mostly to overcome the gravitational well of the star, instead of accelerating the wind to its asymptotic speed. Therefore most of the wind driving energy is injected in the subsonic regions of the flow, since the gravitational potential falls off as the inverse of the distance to the centre of the star. Furthermore, the energy must be added in the form of work done by a force rather than in the form of heat, since radiative cooling is very efficient in the inner parts of the dense winds \citep{1985ASSL..117..229H, 1999isw..book.....L}.

In addition, mass-loss rates in AGB stars increase with pulsation period, bolometric luminosity and optical depth at infrared wavelengths \citep{2010ApJ...719.1274S,2015A&A...581A..60D,2018MNRAS.481.4984M}. 
From these results it is thought that AGB winds are driven by a combination of pulsation-induced shock-waves and radiation pressure on dust \citep{2010A&A...514A..35N}.
These shock-waves eject material into the stellar atmosphere to distances where dust is formed under low temperature and high density conditions. The dust is then accelerated by absorption and scattering of the stellar radiation field and transfers its momentum to the gas through collisions, inducing a slow wind \citep{2018A&ARv..26....1H}.
This is basically the scenario that we include in this study. 

We perform numerical simulations of the stellar wind in both detached and semi-detached binary stars. 
Our goal is to obtain analytical formulae describing the density and outflow velocity of the wind, which can be useful in modelling the interaction of the CE ejecta with the surrounding  medium and the later formation of bipolar PNe \citep{2018ApJ...860...19G}. 
We focus on the effects of orbital separation and stellar mass ratio on the outflow.
To the best of our knowledge, our work is the first attempt to predict the distribution of density and outflow velocity in the stellar wind, based on hydrodynamic models of interacting binaries. In this first article, we obtain 2-dimensional, azimuthally averaged, analytical fits of the outflow structure. Although our simulations are limited by resolution, the qualitative results give an important first insight of such a complex, 3-D problem.

The numerical setup is described in section 2. The effects of orbital parameters and stellar mass ratio \textit{q} on the outflow structure are described in 
section 3. Discussions of results are presented in section 4 and conclusions are provided in section 5.

\section{Method}

We study the interaction of an AGB wind with a companion star using 3-dimensional, hydrodynamical 
simulations.  The computations use a nested mesh scheme. %an adaptive-mesh refinement scheme
We have used stellar pulsations and radiation pressure on dust grains to drive the AGB wind. 
The details are described in the next subsections.

\subsection{The binary system}

%GGS he cambiando un poco esto, porque de hecho, no has usado BEC

The binary system consists of an AGB star with a mass of $M_1=2.2~\textrm{M}_{\sun}$ and a secondary star on the main sequence 
with a mass of $M_2=0.8~\textrm{M}_{\sun}$ (Figure~\ref{colat}).
These values are taken from \citet{2016ApJ...823..142G}. 
We assume a fundamental-mode pulsation period of $P=300~\textrm{d}$ for our AGB star. This period is within the range of values derived from infrared observations \citep{2016ApJ...817..115C}. With the above values of mass and pulsation period, our AGB star has a radius of $R_1=316~\textrm{R}_{\sun}$ \citep{1986ApJ...311..864O}. We adopt an orbital separation of $a=674.3~\textrm{R}_{\sun}$. This choice for the orbital separation ensures that the secondary star lies in a region where dust is absent in the outflow. Observations show that dust is formed at about two stellar radii \citep{2012Natur.484..220N} from the centre of the AGB star.

%{\bf The binary system consist of an AGB star with a mass of $M_1=2.2\,\textrm{M}_{\sun}$ and a secondary star on the main sequence 
%with a mass of $M_2=0.8\,\textrm{M}_{\sun}$ (Figure~\ref{colat}).
%These values are taken from \citet{2016ApJ...823..142G}. The radius of the AGB is $R_1=316\,\textrm{R}_{\sun}$. We adopt an orbital separation of $a=674.3\,\textrm{R}_{\sun}$.}
%{\bf This choice for the orbital separation ensures that the secondary star lies in a region where dust is absent in the outflow. Observations show that dust is formed after two stellar radii \citep{2012Natur.484..220N}.} 
%{\bf With the above values of mass and radius, an AGB star has a fundamental-mode pulsation period $P=300\,\textrm{d}$ \citep{1986ApJ...311..864O}. This period is within the range of values derived from infrared observations \citep{2016ApJ...817..115C}.} 
 
According to the Eggleton formula \citep{1983ApJ...268..368E},

\begin{equation}
\frac{R_{RL}}{a}=\frac{0.49~q_E^{2/3}}{0.6~q_E^{2/3}+ln(1+q_E^{1/3})},
\label{eqrl}
\end{equation}

\noindent with a stellar mass ratio $q_E=M_1/M_2 = 2.75$, the radius of the Roche lobe $R_{RL}$ for the primary star is 0.468 times the orbital separation \textit{a}. Given that $a=674.3~\textrm{R}_{\sun}$, $R_{RL} = 315.6~\textrm{R}_{\sun}$, the primary fills its Roche lobe \citep{1971ARA&A...9..183P}. 

The choice of the stellar mass for the primary in \citet{2016ApJ...823..142G} was based on the fact that progenitor stars of bipolar PNe seems to be more massive than $2~\textrm{M}_{\sun}$. This conclusion is supported by the distribution of bipolar PNe towards lower latitudes above the Galactic plane \citep{1995A&A...293..871C} and their nitrogen enrichment compared to round or elliptical PNe \citep{2006ApJ...651..898S}. 
The choice of the stellar mass for the secondary was based on observational studies which show that the mass ratio distribution $q=M_2/M_1$ for nearby binary stars ($\leq 100~$au) has a maximum close to $\sim0.3$ \citep{2016AJ....152...40G}.

We also increase the mass of the companion to $M_2=1.6~\textrm{M}_{\sun}$, $M_2=2~\textrm{M}_{\sun}$ and $M_2=2.2~\textrm{M}_{\sun}$. These values are chosen to test whether systems with larger mass ratios $q=M_2/M_1$ form gravitationally-bound circumbinary rings \citep{1979ApJ...229..223S, 2016MNRAS.455.4351P} when radiation pressure on dust is taken into account.

The binary system is assumed to follow circular orbits, have synchronous rotation and their spin axes are perpendicular to the orbital plane.

\subsection{The hydrodynamic code}

We use the adaptive-mesh refinement hydrodynamic code \textsc{walicxe3d} \citep{2014MNRAS.437..898T}, which is a 3D extension of the code \textsc{walicxe} \citep{2010ApJ...725.1466E}, to calculate the interaction between the secondary star and the AGB wind.
We approximate the stars as point sources of gravity and neglect self-gravity \citep[e.g.][]{2009ApJ...700.1148D, 2017MNRAS.468.4465C}.

%GGS He introducido X Y Z  y plano XY 

The physical dimensions of the computational domain, in Cartesian XYZ coordinates, are $9,000 \times 9,000 \times 4,500~ \textrm{R}_{\sun}$ ($41.76 \times 41.76 \times 20.88~$au).
This numerical domain consists of three nested meshes fixed to an inertial frame of reference, whose origin coincides with the centre of mass (barycentre) of the system. We prefer to use a nested mesh to concentrate the highest resolution always at the centre of the domain, instead of using the adaptive mesh refinement algorithm.
Each mesh has $128 \times 128 \times 64$ cubical cells, where the shortest dimension 
%GGS
Z is perpendicular to the orbital plane 
%GGS
(XY plane), and each of them is a factor 2 smaller in each axis compared to the next coarser mesh. 
The formal resolution limit in grid-based codes is set by the Nyquist wavelength, which is the lowest wavelength for a wave to reliably propagate across the grid. The Nyquist wavelenght is equal to twice the smallest space interval on the grid \citep{2007nmai.conf.....B}. Based on this, the resolution in our simulations is $\sim 35.2~\textrm{R}_{\sun}$, which corresponds to twice the size of each cell in the innermost mesh ($\sim 17.6~\textrm{R}_{\sun}$).
The orbital plane has a reflective boundary condition, while the other boundaries have free-outflowing conditions.

We consider an ideal gas composed of neutral atoms and assume that the hydrogen to helium number ratio is 10 to 1. Therefore the gas has a ratio of specific heats $\gamma=5/3$ and a constant mean molecular weight per particle $\mu=1.27$.
The radiative cooling is given by vibrational transitions of CO molecules \citep[Appendix 3]{2013pccd.book.....G} and collisional excitation and ionization of hydrogen \citep{1981MNRAS.197..553B}, using the first-order implicit scheme described in \citet{2009ApJS..181..391T}.

\subsection{Numerical approach for the primary star}

For simplifying the hydrodynamic models, we assume a spherical AGB star (i.e. the surface of the primary star is not an equipotential surface).
As an initial condition, the slow wind is setup spherically symmetric with respect to the centre of the AGB star and follows a velocity law of the form \citep[chap. 2]{1999isw..book.....L}

\begin{equation}
v(r_1)=v_0+(v_{\infty}-v_0)(1-R_1/r_1)^\beta,
\label{eq2}
\end{equation}

\noindent where $v(r_1)$ is the wind velocity at a distance $r_1$ from the centre of the AGB star, whose radius is $R_1$. %, whose radius is $R_*$. 
The parameter $\beta$ describes the acceleration in the inner regions of the wind. In cool giant stars the wind accelerates more slowly, corresponding to values of $\beta \ge 1$. We take $\beta=2.0$ but larger values have been used in literature \citep[see also e.g. \citealt{2018ApJ...869....1R}, for warmer giant stars]{2014A&A...570A..67K, 2015A&A...581A..60D}. The outflow velocity in AGB stars is typically 5 to $25~$km\,s$^{-1}$ \citep{2009A&A...499..515R}, for this reason we assign a terminal velocity of $v_{\infty}=10~$km\,s$^{-1}$ and assume that the gas escapes from the stellar surface with a velocity $v_0=0.1~$km\,s$^{-1}$. 
Higher values of the surface velocity are computed in section 3.2.
%Higher values of the surface velocity, \textbf{such as 1.0\,km\,s$^{-1}$ or 5.0\,km\,s$^{-1}$,} at the onset of simulations do not affect substantially our results \textbf{(see Table~\ref{Table3})}.  
The initial density of the background gas is obtained using both the velocity field described in equation~\ref{eq2} and the initial mass-loss rate $\dot{M}$ of the AGB star in the continuity equation. The \citet{1993ApJ...413..641V} parametrization

\begin{equation}
\textrm{log}~\dot{M}(M_{\sun}\,yr^{-1})=-11.4+0.0123~\textrm{P}~ \textrm{(days)}
\end{equation}
 
\noindent gives $\dot{M}\sim2\times10^{-8}~\textrm{M}_{\sun}\,\textrm{yr}^{-1}$, for the adopted period of pulsation $P=300~\textrm{d}$.
%GGS This value for \textit{P} is selected, based on results obtained with the BEC model 
%GGS(see section 3.1) 
%GGS Furthermore, 
%GGS Esto esta ya mas arriba: This period is within the range of values derived from infrared observations \citep{2016ApJ...817..115C}.} 

The initial temperature field of the gas $T_\textrm{eq}$ at a distance $r_1$ and $r_2$ from the centre of the primary and secondary star, respectively, is \citep{1986Ap&SS.124....5D}

\begin{equation}
T_\textrm{eq} = \left(\frac{1}{2}\right)^{\frac{1}{4}}\left(\left[ 1-\sqrt{1-\left(\frac{R_1}{r_1}\right)^{2}}\right] T_1^4  +  \left[ 1-\sqrt{1-\left(\frac{r_s}{r_2}\right)^{2}}\right] T_s^4 \right)^{\frac{1}{4}},
\end{equation}

We assume an effective temperature of $T_1=3000~\textrm{K}$ for the AGB star. This is a common temperature for Mira variable stars pulsating in their fundamental mode \citep{2002AJ....124.1706V}.

\noindent We adopt an effective temperature $T_s=5000~\textrm{K}$ and a stellar radius $r_s=1~R_{\sun}$ for our secondary star of mass $M_2=0.8~\textrm{M}_{\sun}$, following the temperature-mass and mass-radius diagrams from \citet{2010A&ARv..18...67T}.

The wind from the AGB is driven by stellar pulsations coupled with radiation pressure on dust
%GGS
grains. Simple harmonic variations of the gas velocity $V(t)$ at the position $R(t)$ are employed to model the pulsations as follows

\begin{equation}
V(t)=\Delta u~cos(2\pi t/P),
\end{equation}

\begin{equation}
R(t)=R_0+\frac{\Delta u}{2\pi/P}~sin(2\pi t/P),
\end{equation}

\noindent We assume an initial position of pulsation of $R_0=0.9~R_1$. This value is chosen because  pulsations in AGB stars are believed to be driven in the hydrogen-ionization zone, located approximately at $0.1-0.2~R_1$ below the photosphere \citep{2003A&A...397..943B, 2016AstL...42..665F}. 
$\Delta u=5~$km\,s$^{-1}$ and $P=300~\textrm{d}$ are the constant velocity amplitude and period of the pulsations, respectively. 
The former value is chosen from the range of radial velocities in AGB stars derived from observed Doppler-shifts of spectral lines \citep{2002ASPC..259..556L, 2010A&A...514A..35N}.

From the equations of equilibrium \citep[chap. 9]{1988ApJ...329..803A, 1967aits.book.....C}, the temperature and density structure of the AGB surface layers are given by

\begin{equation}
T(r_1)=T_1+\frac{1.79}{2.18}\frac{G\,M_1\,\mu\,m_\textrm{H}}{k_\textrm{B}}\left(\frac{1}{r_1}-\frac{1}{R_1}\right),
\label{eq3}
\end{equation}
\begin{equation}
\rho(r)=\left[\left(\frac{1.79}{2.18}\right)\left(\frac{16\,\pi\,G\,M_1\,a\,c\,\mu\,m_\textrm{H}}{3\,k_\textrm{B}\,L_1\,\kappa }\right)\right]T^{3}(r_1),
\label{eq4}
\end{equation}

\noindent where $k_\textrm{B}$, \textit{G}, $m_\textrm{H}$, \textit{a}, \textit{c} and $L_1$ are the Boltzmann constant, the gravitational constant, the atomic mass unit, the radiation constant, the speed of light and the AGB star luminosity, respectively. 
Note that the law of opacity $\kappa$ is given by the negative hydrogen ion $\textrm{H}^-$ \citep[chap. 7]{2013sepp.book.....I}.
Although we have the AGB stellar structure from \citet{2016ApJ...823..142G}, it is preferable in our case to use the analytical expressions as in equations~\ref{eq3} and \ref{eq4}, since the grid resolution is limited, and the interpolation produces more numerical errors than the analytical formulae. The remaining AGB interior is assumed to be homogeneous and isothermal.

We have considered the AGB star as a point source of radiation, thus the radiative acceleration takes the form:

\begin{equation}
a_\textrm{rad}=\frac{k_\textrm{D} L_1}{4\pi r_1^{2}c} e_r ,
\label{eq7}
\end{equation}

\noindent where $e_r$ is the unit radial vector and $k_\textrm{D}$ is the dust opacity, which is approximated as \citep{1988ApJ...329..299B}

\begin{equation}
k_\textrm{D}=\frac{k_\textrm{max}}{1+\textrm{e}^{(T_\textrm{eq}-T_\textrm{cond})/\xi}},
\label{eq8}
\end{equation}

\noindent $T_\textrm{cond}=1500~K$ is the condensation temperature and its range spreads over $\xi=200~K$. These values are selected because the most abundant dust species in carbon-rich stars, namely amorphous carbon and silicon carbide \citep{2015A&A...583A.106R}, have condensation temperatures between 1300 and 1700 K, approximately \citep{1995Metic..30..661L,2008ApJ...681.1557L}.
The parameter $k_\textrm{max}$, which is the maximum value of dust opacity, is a constant adjusted to give slow wind expansion velocities of $\sim10-20~$km\,s$^{-1}$ towards the pole. The AGB luminosity $L_1$ is calculated from the stellar effective temperature, radius and pulsation properties 

\begin{equation}
L_1=\pi ac T_1^4 \left(R_1+\frac{\Delta u}{2\pi/P}sin(2\pi t/P)\right)^2 , 
\label{eq5}
\end{equation}

\subsection{Numerical approach for the secondary star}

In our simulations the secondary star is not resolved; it is a point source, since its radius (equal to one solar radius) is smaller than a grid cell of our computational mesh. Therefore,  its gravitational potential would induce extremely large accelerations to the nearby gas and, consequently, infinitesimal time-steps. For avoiding this, we smoothed the potential following \citet{1993A&A...280..141R}

\begin{equation}
\phi(r_2)=\frac{-GM_2}{\sqrt{r_2^2+\epsilon^2 \delta^2 \textrm{e}^{-(r_2/\epsilon \delta)^2}}},
\label{eq1}
\end{equation}

\noindent The smoothing length $\delta$ is equal to the size of one cell in the innermost mesh and $r_2$ is the distance to the secondary star of mass $M_2$. Following \citet{2016MNRAS.455.3511S} the parameter $\epsilon$ is equal to 3, which ensures a finite gravitational potential at the secondary star position. With this value for $\epsilon$, the smoothed potential is identical to the true potential at $\sim7$ cell widths from the secondary star (see discussion).
 
Lower values for $\epsilon$ produce similar results at the cost of smaller time-steps, which increases the total computational time.

\section{Results}

\subsection{3-D numerical results}

%GGS esto lo puse aqui
The simulations are first run during a couple of orbits until they reached a steady-state configuration.
After that, we calculate 40 orbital cycles to make a time average of the density and outflow velocity of the stellar wind, in order to obtain an axisymmetric, 2-dimensional structure.

Table~\ref{table1} shows different input parameters such as orbital separation \textit{a}, mass of the stellar components $M_1$ and $M_2$, mass ratio \textit{q}, orbital period $P_{orb}$ and maximum dust opacity $k_{max}$ for each binary system. The Model 1 (our fiducial model) is shown in the first line.

%GGS 
The computed mass-loss rates of each model are shown in Table~\ref{Table1a}.
The mass-loss rates are calculated through a spherical surface of radius $4300~\textrm{R}_{\sun}$ ($20.0~$au), close to the outer boundaries of the simulation, where the out-flowing material is supersonic.

Figure~\ref{fig1} shows a density snapshot of Model 1 at the orbital plane (XY plane) and Figure~\ref{fig2} shows its time-averaged density (top panel) and outflow velocity (bottom panel) perpendicular to the XZ plane. We find that mass loss in the binary star occurs mainly through the outer Lagrangian point $L_2$. The resultant outflow develops into a spiral at low distances from the binary, that is quickly smoothed out by shocks and becomes an excretion disk at larger distances from the stars. This leads to the formation of an outflow structure with an equatorial density enhancement, i.e., with a large pole-to-equator density ratio, as also shown in \citet{2017MNRAS.468.4465C}.

Figure~\ref{fig3} shows the time-averaged density (top panel) and outflow velocity (bottom panel) profiles, as functions of the coordinate X. The coordinate X indicate the distance from the centre of mass.
These profiles are obtained at the equatorial plane to study mass-loss sensitivity to orbital separation. 
Far from the barycentre, the gravitational potential of the binary system approximates that of a single star. Consequently, at distances larger than $\sim8~$au the outflow in these models follows similar power-law density and velocity distributions (see subsection 3.2). %$\sim2000\,\textrm{R}_{\sun}$

\begin{figure}
	\centering
	\includegraphics[width=1.0\columnwidth]{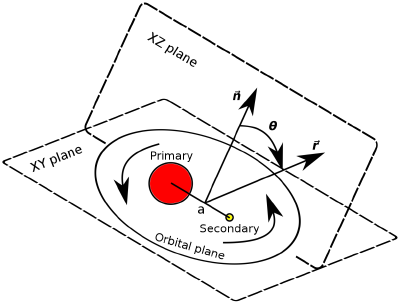}
	\caption{Colatitude angle $\theta$ between the position vector $\vec{r}$ of each mesh cell and the unit normal vector $\vec{n}$ to the orbital plane. The orbital separation \textit{a} is indicated by the line segment joining the stars.}
	\label{colat}
\end{figure}

\begin{figure}
	\centering
	\includegraphics[width=1.0\columnwidth]{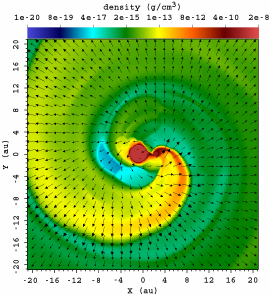}
	\caption{Density map in logarithmic scale at the XY plane for our fiducial model (Model 1). Black arrows represent the velocity field. The centre of mass (barycentre) is located at ($0$, $0$).} %($4500\,\textrm{R}_{\sun}$, $4500\,\textrm{R}_{\sun}$)
	\label{fig1}
\end{figure}

\begin{figure}
	\centering
	\includegraphics[width=1.0\columnwidth]{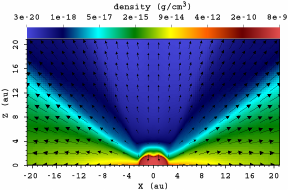}\vspace{1mm}
	\includegraphics[width=1.0\columnwidth,left]{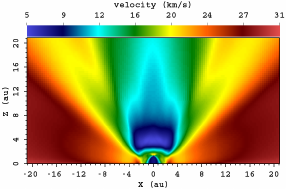} 
	\caption{Time-averaged density (top) and outflow velocity (bottom) maps in logarithmic and linear scale, respectively, for our fiducial model at the XZ plane. Black arrows represent the velocity field.}
	\label{fig2} 

\end{figure}

\begin{figure}
	\centering
	%\subfloat[]{\includegraphics[width=0.5\textwidth]{Plot1.png}\label{fig1}}%1
	\includegraphics[width=1.0\columnwidth]{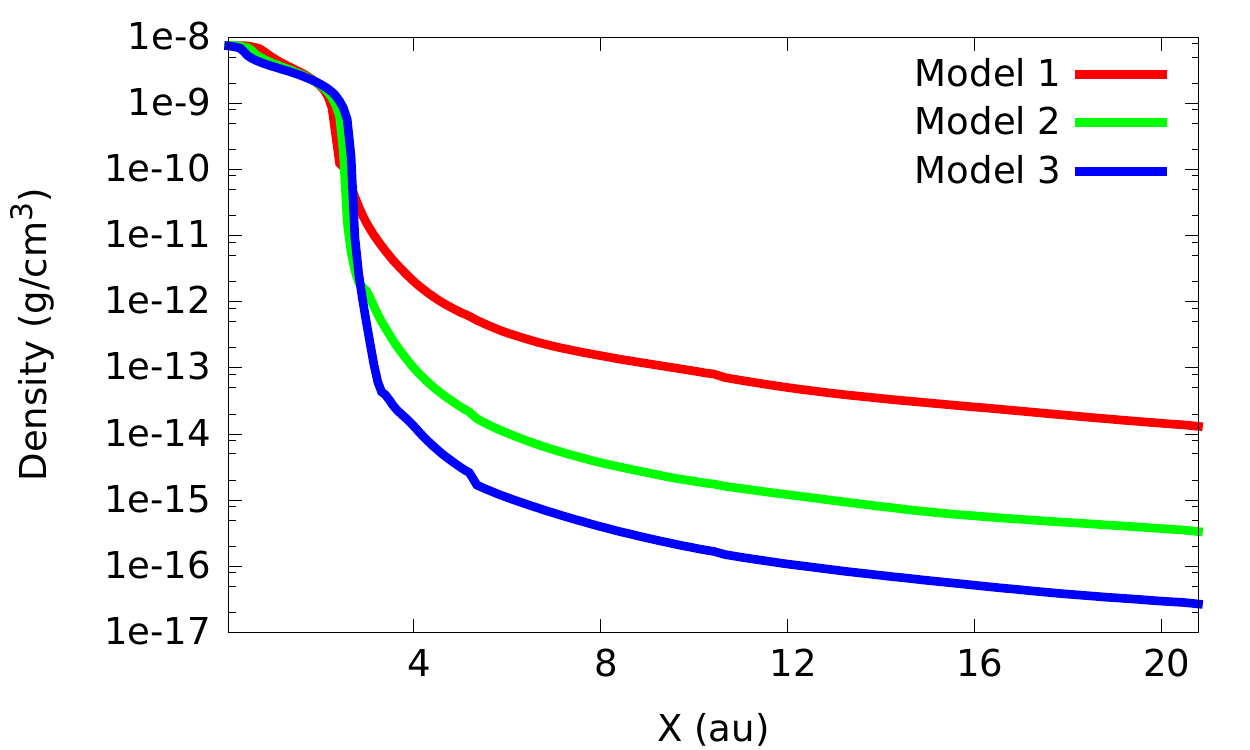} %\label{fig2}%2
	%\subfloat[]{\includegraphics[width=0.5\textwidth]{Plot3.png}\label{fig3}}%3
	\hfill
	\includegraphics[width=1.0\columnwidth]{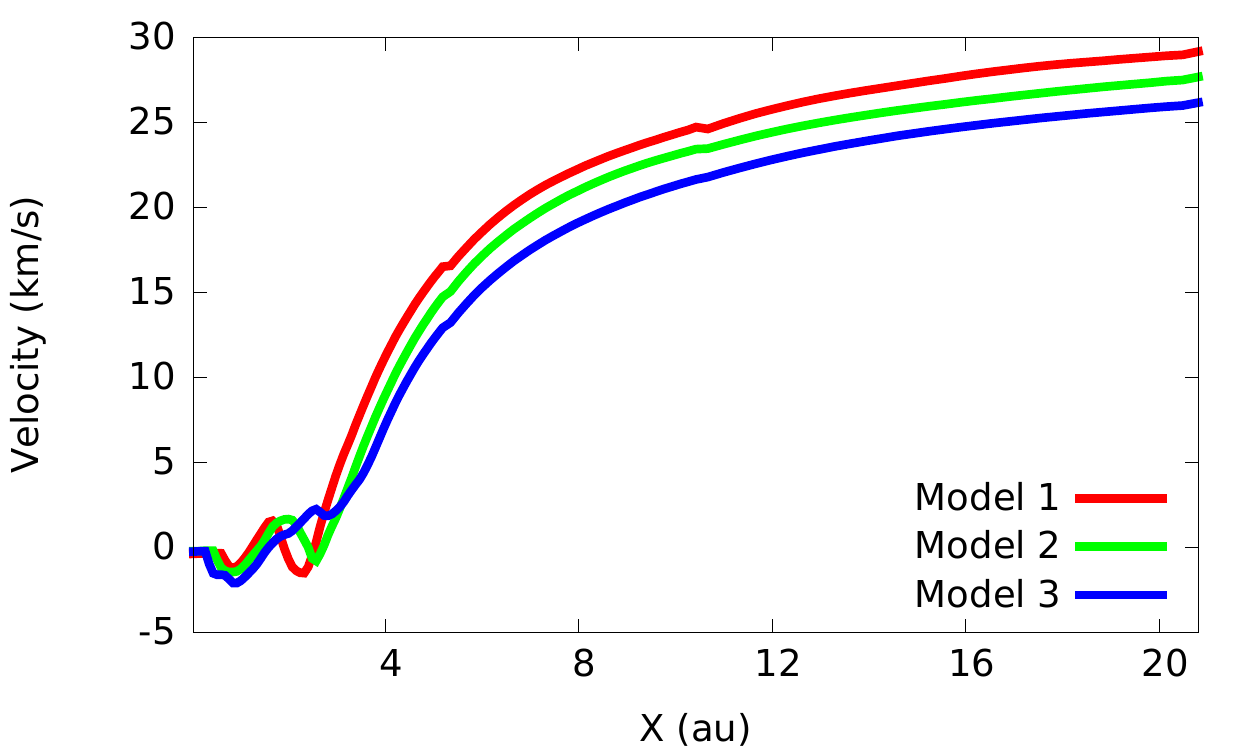} %\label{fig4}%4
	%\\
	\caption{Time-averaged density (top) and outflow velocity (bottom). The horizontal axes indicate the distance from the centre of mass.}
	\label{fig3}
\end{figure}

\begin{figure}
	\centering
	\includegraphics[width=1.0\columnwidth]{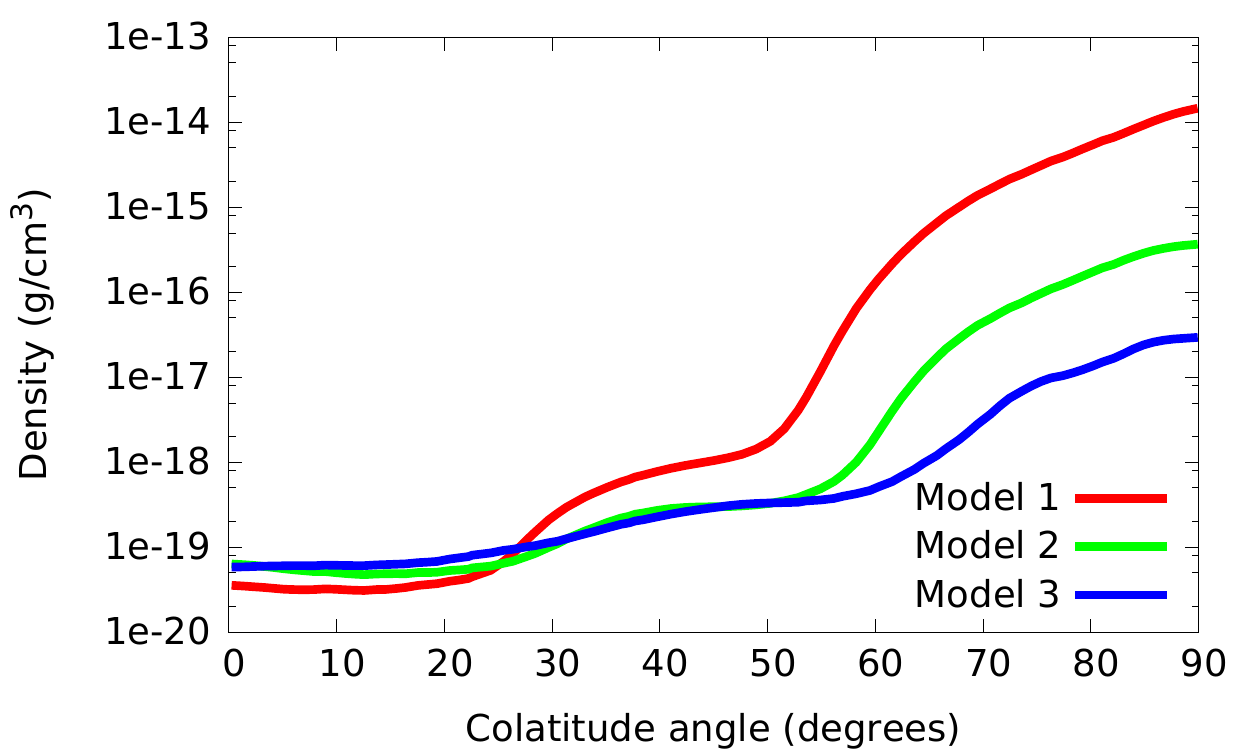}
	\hfill
	\includegraphics[width=1.0\columnwidth]{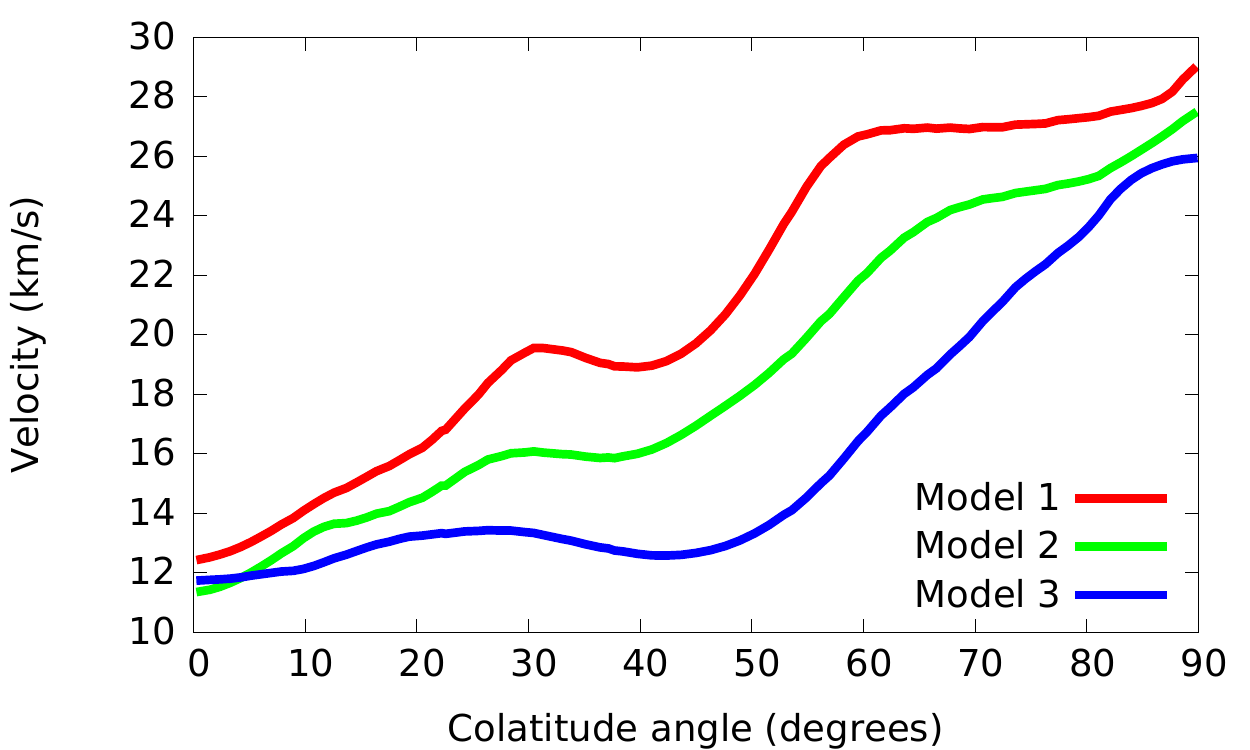}
	\caption{Time-averaged density (top) and outflow velocity (bottom) at $20.0~$au from the barycentre of the system for Models 1, 2 and 3. The horizontal axes indicate the colatitude angle which goes from $0\degr$ at the pole to $90\degr$ at the XY plane.} %$4300\,\textrm{R}_{\sun}$
	\label{fig4}
\end{figure}

%GGS
Figure~\ref{fig4} shows the time-averaged density (top panel) and outflow velocity (bottom panel) profiles, as functions of the colatitude angle. These profiles are obtained at equidistant points from the barycentre of the system, in the outermost mesh. 
This figure shows the pole-to-equator density ratio, which reaches a maximum value of $\sim10^5$ at Roche-Lobe Overflow (RLOF) state for Model 1.
The colatitude angle goes from $0\degr$ at the pole to $90\degr$ at the XY plane (Figure~\ref{colat}).

Models 1, 2 and 3 have equal stellar masses but different orbital periods and separations.  
When the orbital separation is smaller, the outflow structure changes in the following ways. 
First, at distances from the centre of mass greater than $\sim4~$au, the circumbinary gas is denser in Model 1 with respect to Models 2 and 3 at the orbital plane, as shown in the top panel of Figure~\ref{fig3}. %$\sim1000\,\textrm{R}_{\sun}$
This is caused by a larger mass-loss rate from the binary system through the outer Lagrangian points.
Second, a stronger deviation from spherical symmetry in the mass-loss geometry is produced, or in other words, the pole-to-equator density ratio increases up to $\sim10^5$, as it is shown in the top panels of Figure~\ref{fig2} and \ref{fig4}. 
This is caused by a stronger gravitational focusing of the winds towards the companion star. 
Third, the outflow is faster (bottom panels of Figure~\ref{fig3} and \ref{fig4}). This is a consequence of the momentum imparted to the wind by the secondary star with a faster orbital motion. 
The above results give rise to an increase of the mass-loss rate (Table~\ref{Table1a}).
%Finally, the mass-loss rate increases \textbf{(Table~\ref{Table1a})}. 

\begin{figure}
	\centering
	\includegraphics[width=1.0\columnwidth]{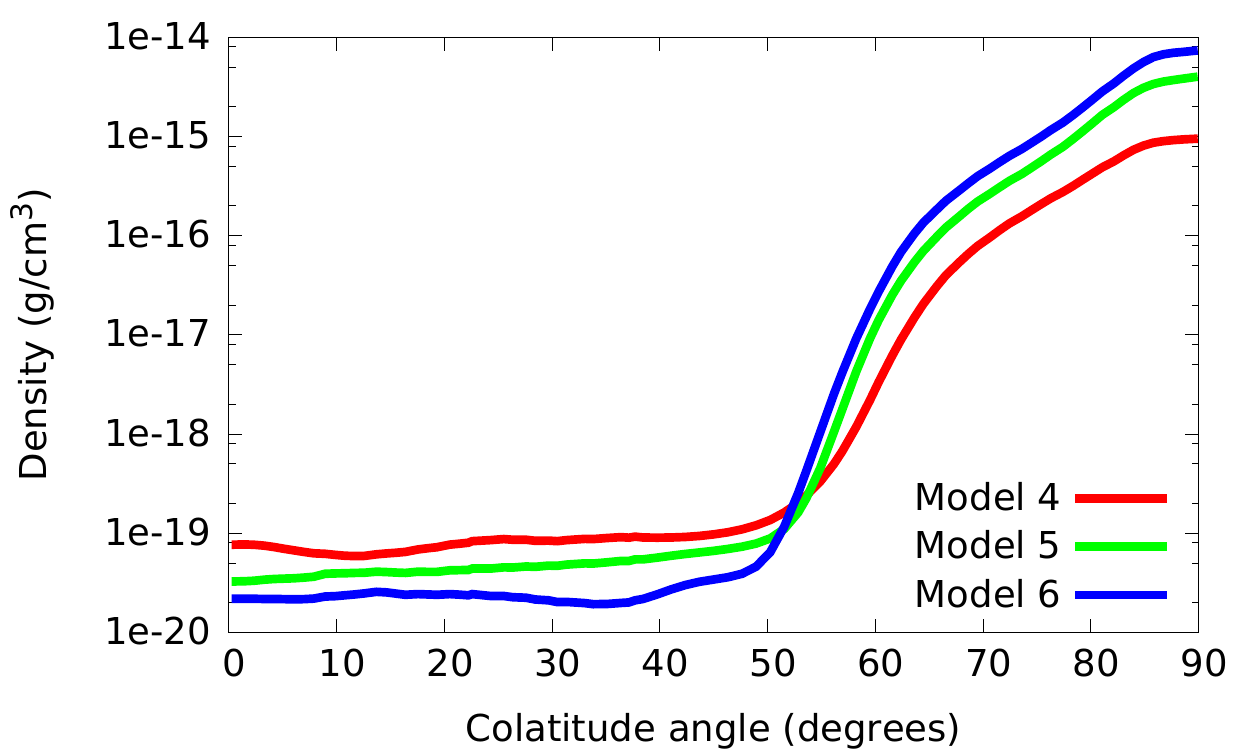}
	\hfill
	\includegraphics[width=1.0\columnwidth]{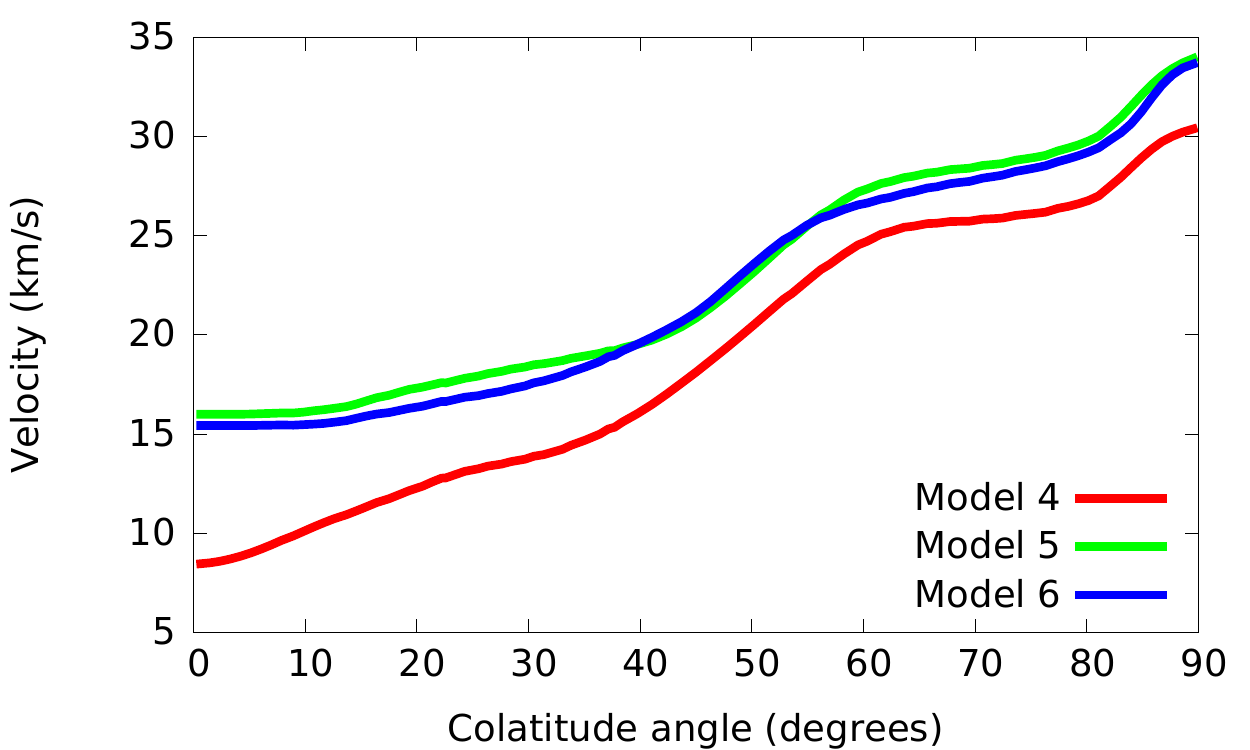}
	\caption{Same as Fig.~\ref{fig4} for models 4, 5 and 6.} 
	\label{fig5}
\end{figure}

Models 4, 5 and 6 have equal orbital separation but different companion stellar mass.  
This produces the following changes in the outflow structure when the mass ratio $q=M_2/M_1$ increases towards unity. First, more material is lost through the outer Lagrangian points and less gas reaches the distant polar regions, as a consequence the pole-to-equator density ratio also increases 
(top panel of Figure~\ref{fig5}). Second, the mass-loss rate is enhanced, as can be seen in Table~\ref{Table1a}.
If both the companion stellar mass and the maximum dust opacity $k_\textrm{max}$ increase, then the outflow velocity also increases as shown in the bottom panel of Figure~\ref{fig5}.   

\begin{figure}
	\centering
	\includegraphics[width=1.0\columnwidth]{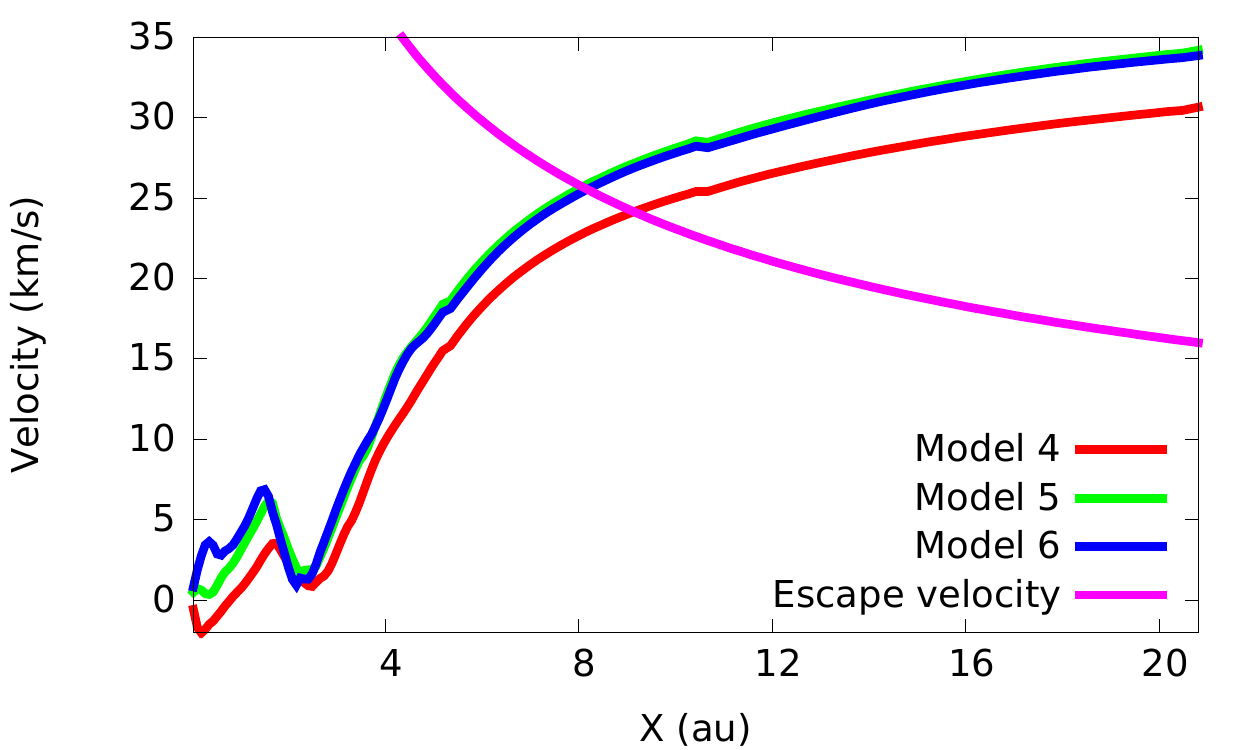}
	\caption{Time-averaged outflow velocity, as function of the X coordinate, for Models 4, 5 and 6. Also shown the escape velocity relative to the barycentre.}
	\label{fig10}
\end{figure}

\begin{figure}
	\centering
	\includegraphics[width=1.0\columnwidth]{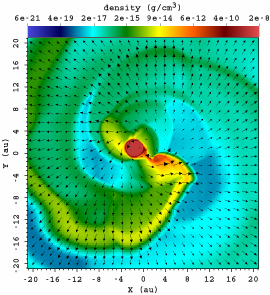}
	\caption{Density map in logarithmic scale at the XY plane for Model 4. Black arrows represent the velocity field. The barycentre is located at ($0$, $0$).} %($4500\,\textrm{R}_{\sun}$, $4500\,\textrm{R}_{\sun}$)
	\label{fig10-1}
\end{figure}

\begin{figure}
	\centering
	\includegraphics[width=1.0\columnwidth]{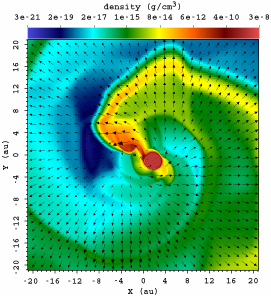}
	\caption{Same as Fig.~\ref{fig10-1} for Model 5.} 
	\label{fig10-2}
\end{figure}

\begin{figure}
	\centering
	\includegraphics[width=1.0\columnwidth]{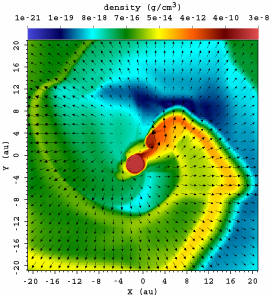}
	\caption{Same as Fig.~\ref{fig10-1} for Model 6.} 
	\label{fig11}
\end{figure}

%$\sim2000\,\textrm{R}_{\sun}$
Beyond $\sim10~$au from the barycentre, the time-averaged outflow velocity at the orbital plane of Models 4, 5 and 6 becomes larger than the escape velocity relative to the barycentre, as shown in Figure~\ref{fig10}. In Models 4, 5 and 6 the stellar mass ratios are equal to 0.73, 0.91 and 1, respectively. Therefore if radiation pressure on dust is included, the gas ejected through the Lagrangian $L_2$ point does not form an outer ring of gravitationally bound material when $q=M_2/M_1=0.73$ (Figure~\ref{fig10-1}) or when $q=M_2/M_1>0.78$,
%GGS: \citep{1979ApJ...229..223S,2016MNRAS.455.4351P}
 as seen in Figures~\ref{fig10-2} and \ref{fig11}.

For comparison, we have made an adiabatic model (Table~\ref{table1}) without radiative cooling. In this model, the gas behind the pulsation-induced shock can no longer radiate its 
internal energy and expands adiabatically before the next shock appears. As a consequence, the gas velocity increases. Because of that, the gravitational focusing of the AGB wind towards the orbital plane is underestimated \citep{2002A&A...385..205G} since there is a larger thermal pressure that works against gas compression.

\subsection{Analytical fits}
Finally, we make a curve fitting to the profiles to find analytical expressions of the outflow structure.
The variation with $\theta$ of density and outflow velocity in the stellar wind is described using several Gaussian functions in equations~\ref{eq9} and \ref{eq10}, respectively.
In these equations, density and outflow velocity are given by the coefficient \textit{s} near the symmetry axis and by $b_i$ at colatitude angle $c_i$. The coefficient $d_i$ represents the width of the Gaussian functions. Table~\ref{table2} shows numerical values of the coefficients for our Models.
Density (g\,cm$^{-3}$) and outflow velocity (km\,s$^{-1}$) as functions of both distance \textit{r} (in au) from the barycentre and colatitude angle $\theta$ (in degrees) are given in equations~\ref{eq11} and \ref{eq12}, respectively. 
In the former equation the dependence on distance is fitted to a power-law function $\rho\,\propto\,r^{-f}$. While in the latter equation, we use a velocity field similar to that described in equation~\ref{eq2}; however we replace the constant surface velocity $v_0$ by equation~\ref{eq10}.

\begin{equation}
\Psi(\theta)=s+\sum_{i=1}^{5} b_i \textrm{e}^{-(\theta-c_{i})^2/d_{i}^{2}},
\label{eq9}
\end{equation}

\begin{equation}
\vartheta(\theta)=s+\sum_{i=1}^{5} b_i \textrm{e}^{-(\theta-c_{i})^2/d_{i}^{2}},
\label{eq10}
\end{equation}

\begin{equation}
\frac{\rho(r,\theta)}{\rho_\textrm{ref}}=\Psi(\theta)\left(\frac{r_c}{r}\right)^f,
\label{eq11}
\end{equation}

\begin{equation}
\frac{v(r,\theta)}{V_\textrm{orb}}=\vartheta(\theta)+(\vartheta(\theta)-v_{\infty})*\left(\frac{r_c}{r}-1\right)^f,
\label{eq12}
\end{equation}

\noindent Given that these expressions could be used as initial conditions characterizing the circumbinary medium in simulations of aspherical PNe at larger spatial dimensions, we choose the length scale parameter $r_c$ as far from the binary star as possible, in a region where the outflow is unbound and its morphology has not changed considerably as compared to inner regions. Taking the above into account, $r_c=20.0~$au. %$r_c=4300\,\textrm{R}_{\sun}$.
The parameter $v{_\infty}$ is the wind terminal velocity resulting from the simulations. $V_\textrm{orb}=(2\pi a)/P_\textrm{orb}$ and $\rho_\textrm{ref}=\dot{M}/(4\pi r_c^2 V_\textrm{orb})$ are, respectively, velocity and density scale factors. The orbital separation \textit{a} and the orbital period P$_\textrm{orb}$ for each model are given in Table~\ref{table1}, while the total mass-loss rates $\dot{M}$ are shown in Table~\ref{Table1a}.

Figure~\ref{plotfit} and Figures~\ref{plotfit2} to \ref{plotfit7} in the appendix show the analytical fit to the time-averaged density and outflow velocity profiles, as functions of the colatitude angle, for our models using equations~(\ref{eq9}) and (\ref{eq10}), respectively.

Finally, Table~\ref{Table3} shows the time-averaged density and outflow velocity of the stellar wind of additional models. For each of these models the initial wind velocity $v_0$ at the surface of the AGB star is different, while the input parameters such as orbital separation, stellar masses and maximum dust opacity are equal to those in our Model 1 (Table 1). Values for density and outflow velocity were taken at a fixed distance of $20.0~$au from the barycentre and at different colatitude angles $\theta$. As can be seen in Table~\ref{Table3}, the outflow hydrodynamical structure is not modified considerably with different values for $v_0$. %$4300\,\textrm{R}_{\sun}$

\begin{table*}
	\caption{Input parameters for the binary models.} 
	\label{table1}
	\tabcolsep=0.11cm
	\begin{tabular}{c c c c c c c c}
		\hline
		
		Models & \textit{a} ($\textrm{R}_{\sun}$) & $M_1$ ($\textrm{M}_{\sun}$) & $M_2$ ($\textrm{M}_{\sun}$) & $q=M_2/M_1$ & $P_{\textrm{orb}}$ (yr) & $k_\textrm{max}$ & Remarks\\  \hline
		{\bf 1} & {\bf 674.3} & {\bf 2.2} & {\bf 0.8} & {\bf 0.36} & {\bf 3.2} & {\bf 5.6} & {\bf Fiducial model in RLOF}\\ 
		2 & 809.2 & 2.2 & 0.8 & 0.36 & 4.2 & 5.6 & \\
		3 & 944.0 & 2.2 & 0.8 & 0.36 & 5.3 & 5.6 & \\
		4 & 944.0 & 2.2 & 1.6 & 0.73 & 4.7 & 6.6 & \\ 
		5 & 944.0 & 2.2 & 2.0 & 0.91 & 4.5 & 8.5 & \\
		6 & 944.0 & 2.2 & 2.2 & 1.0 & 4.4 & 8.8 & \\
		7 & 674.3 & 2.2 & 0.8 & 0.36 & 3.2 & 5.6 & Adiabatic simulation\\ 
		\hline
	\end{tabular}
	%\end{changemargin}
\end{table*}

\begin{table*}
	\caption{Mass-loss rates for the binary models resulting from our calculations.}
	\label{Table1a}
	\tabcolsep=0.11cm
	\begin{tabular}{c c}
		\hline
		Models & $\dot{M}~(\textrm{M}_{\sun}\,yr^{-1})$\\ \hline
		1 & 5.8e-5 \\
		2 & 1.6e-6 \\
		3 & 1.2e-7 \\
		4 & 4.4e-6 \\
		5 & 1.8e-5 \\
		6 & 3.2e-5 \\
		7 & 7.2e-4 \\
		\hline
	\end{tabular}
\end{table*}

%\begin{sidewaystable}
\begin{table*}
	%\begin{changemargin}{-2cm}{-2cm}
	\caption{Coefficients for the analytical fits of the outflow structure} 
	\label{table2}
	%   \setlength{\tabnotewidth}{0.95\linewidth}
	%  \setlength{\tabcolsep}{0.8\tabcolsep} \tablecols{13}
	%\resizebox{\textwidth}{!}
	%{
	\tabcolsep=0.11cm
	\begin{tabular}{c c c c c c c c c c c c c c c c c c}  %\toprule 
		\hline
		Function & \textit{s} & $b_1$ & $c_1$ & $d_1$ & $b_2$ & $c_2$ & $d_2$ & $b_3$ & $c_3$ & $d_3$ & $b_4$ & $c_4$ & $d_4$ & $b_5$ & $c_5$ & $d_5$ & $f$\\% \midrule
		\hline
		%\multirow{2}{*}{1} 
		$\rho_1(r,\theta)$ & $2.9e-5$ & $5.4e-4$ & $39$ & $8.2$ & $8e-4$ & $49$ & $7$ & $1.1$ & $72$ & $7.9$ & $3.5$ & $80.5$ & $6.5$ & $11.6$ & $90$ & $6.2$ & $2.55$ \\
		$v_1(r,\theta)$ & $0.42$ & $5.9e-2$ & $13.3$ & $7.8$ & $0.2$ & $30.4$ & $11.4$ & $0.21$ & $57.8$ & $13.3$ & $0$ & $0$ & $0$ & $0.54$ & $90$ & $37.3$ & $1$ \\
		%\multirow{2}{*}{2}
		\hline
		$\rho_2(r,\theta)$ & $1.5e-3$ & $4.1e-4$ & $0$ & $6$ & $7.1e-3$ & $42.5$ & $10.5$ & $8.8e-3$ & $58.5$ & $9$ & $1.48$ & $74.5$ & $7.9$ & $10.7$ & $90$ & $10.2$ & $2.55$ \\
		$v_2(r,\theta)$ & $0.42$ & $5.2e-2$ & $11.4$ & $6.5$ & $0.16$ & $28.1$ & $13$ & $0.25$ & $54.3$ & $15.5$ & $0.23$ & $68.2$ & $11.1$ & $0.58$ & $90$ & $18.6$ &  $1$ \\
		%\multirow{2}{*}{3}
		\hline
		$\rho_3(r,\theta)$ & $2.2e-2$ & $5.1e-3$ & $27$ & $7$ & $9.2e-2$ & $49$ & $14.5$ & $0.15$ & $67$ & $10$ & $3.7$ & $79$ & $8$ & $10.1$ & $90$ & $7$ & $2.5$\\
		$v_3(r,\theta)$ & $0.47$ & $7.7e-2$ & $26.1$ & $16.1$ & $0$ & $0$ &$0$ &$0$ &$0$ &$0$ & $0.2$ & $66.2$ & $14.4$ & $0.57$& $90$ & $17.9$ &$1$ \\ %\bottomrule
		\hline 
		$\rho_4(r,\theta)$ & $5e-4$ & $3.2e-4$ & $0$ & $8$ & $5.1e-4$ & $35$ & $22$ & $1.8e-3$ & $58$ & $8.3$ & $0.62$ & $71.6$ & $6.4$ & $10.8$ & $90$ & $10.9$ & $2.45$ \\
		$v_4(r,\theta)$ & $0.24$ & $0.16$ & $22.4$ & $19.9$ & $0$ & $0$ & $0$ & $0.67$ & $66.5$ & $27.2$ & $8.2e-2$ & $78.1$ & $6.4$ & $0.5$ & $90$ & $8.2$ & $1$ \\
		\hline 
		$\rho_5(r,\theta)$ & $9.5e-5$ & $0$ & $0$ & $0$ & $4.3e-5$ & $30$ & $15.9$ & $1.2e-4$ & $50$ & $8.9$ & $0.63$ & $72.1$ & $6.9$ & $11.2$ & $90$ & $9.5$ & $2.6$ \\
		$v_5(r,\theta)$ & $0.54$ & $0$ & $0$ & $0$ & $6.9e-2$ & $29$ & $14.5$ & $0.41$ & $64.5$ & $20$ & $0.18$ & $80$ & $9.5$ & $0.45$ & $90$ & $7.5$ & $1$ \\
		\hline 
		$\rho_6(r,\theta)$ & $3.2e-5$ & $9.9e-6$ & $17$ & $10$ & $0$ & $0$ & $0$ & $3.6e-5$ & $51$ & $8$ & $0.66$ & $72$ & $7$ & $12.1$ & $90$ & $9.3$ & $2.45$ \\
		$v_6(r,\theta)$ & $0.52$ & $0$ & $0$ & $0$ & $6.7e-2$ & $34$ & $16.2$ & $0.19$ & $55.7$ & $14$ & $0.46$ & $83.2$ & $26.1$ & $0.19$ & $90$ & $5.6$ & $1$ \\
		\hline
		$\rho_7(r,\theta)$ & $0.2$ & $2.9e-2$ & $20$ & $15$ & $0$ & $0$ & $0$ & $0.33$ & $60$ & $20$ & $0.43$ & $71$ & $10$ & $9.4$ & $90$ & $10.5$ & $2.1$ \\
		$v_7(r,\theta)$ & $0.7$ & $1.7e-2$ & $0$ & $7$ & $9.2e-2$ & $39$ & $16$ & $0$ & $0$ & $0$ & $6.7e-2$ & $65$ & $17$ & $0.27$ & $90$ & $12$ & $1$ \\
		\hline
	\end{tabular}
	%}
	%\end{changemargin}
\end{table*}
%\end{sidewaystable}

\begin{table*}
	\caption{Density and outflow velocity of the stellar wind for models with different initial wind velocities at the surface of the AGB star.}
	\label{Table3}
	\tabcolsep=0.2cm
	\begin{tabular}{c c c c c c c}
	\hline
	                    & \multicolumn{2}{|c|}{Model with $v_0=0.1~$km\,s$^{-1}$} & \multicolumn{2}{|c|}{Model with $v_0=1.0~$km\,s$^{-1}$} & \multicolumn{2}{|c|}{Model with $v_0=5.0~$km\,s$^{-1}$}\\
	\hline                 
	Colatitude angle $\theta$ & $\rho$ (g\,cm$^{-3})$ & $v$ (km\,s$^{-1})$ & $\rho$ (g\,cm$^{-3})$ & $v$ (km\,s$^{-1})$ & $\rho$ (g\,cm$^{-3})$ & $v$ (km\,s$^{-1})$\\
	\hline
	5.2                 & 4.5e-20               & 10.9               & 4.4e-20               & 10.9               & 4.2e-20               & 11.0              \\
	14.5                & 4.3e-20               & 15.3               & 4.2e-20               & 15.4               & 3.8e-20               & 15.5              \\
	24.5                & 6.1e-20               & 18.1               & 6.1e-20               & 18.1               & 5.7e-20               & 18.3              \\
	34.8                & 4.1e-19               & 19.0               & 4.1e-19               & 19.0               & 4.0e-19               & 19.2              \\
	45                  & 2.0e-18               & 17.9               & 2.0e-18               & 17.9               & 1.9e-18               & 18.1              \\
	55.2                & 2.9e-17               & 23.8               & 2.9e-17               & 23.8               & 2.9e-17               & 23.8              \\
	64.7                & 1.3e-15               & 26.8               & 1.3e-15               & 26.8               & 1.3e-15               & 26.8              \\
	75.6                & 2.5e-14               & 28.7               & 2.5e-14               & 28.7               & 2.5e-14               & 28.7              \\
	85.1                & 8.0e-14               & 29.1               & 8.0e-14               & 29.1               & 8.0e-14               & 29.2              \\
	\hline
	\end{tabular}
\end{table*}

\begin{figure}
	\centering
	\includegraphics[width=1.0\columnwidth]{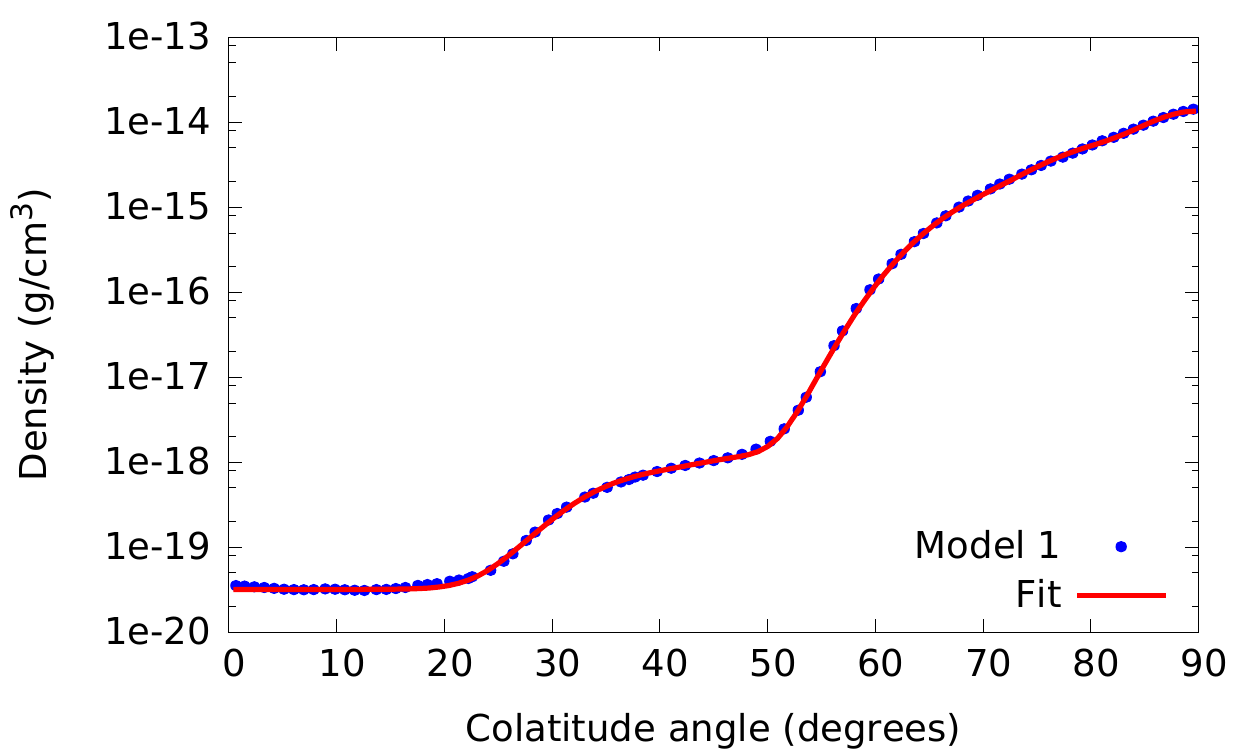} %\label{fit} fit-rho-model1
	\hfill
	\includegraphics[width=1.0\columnwidth]{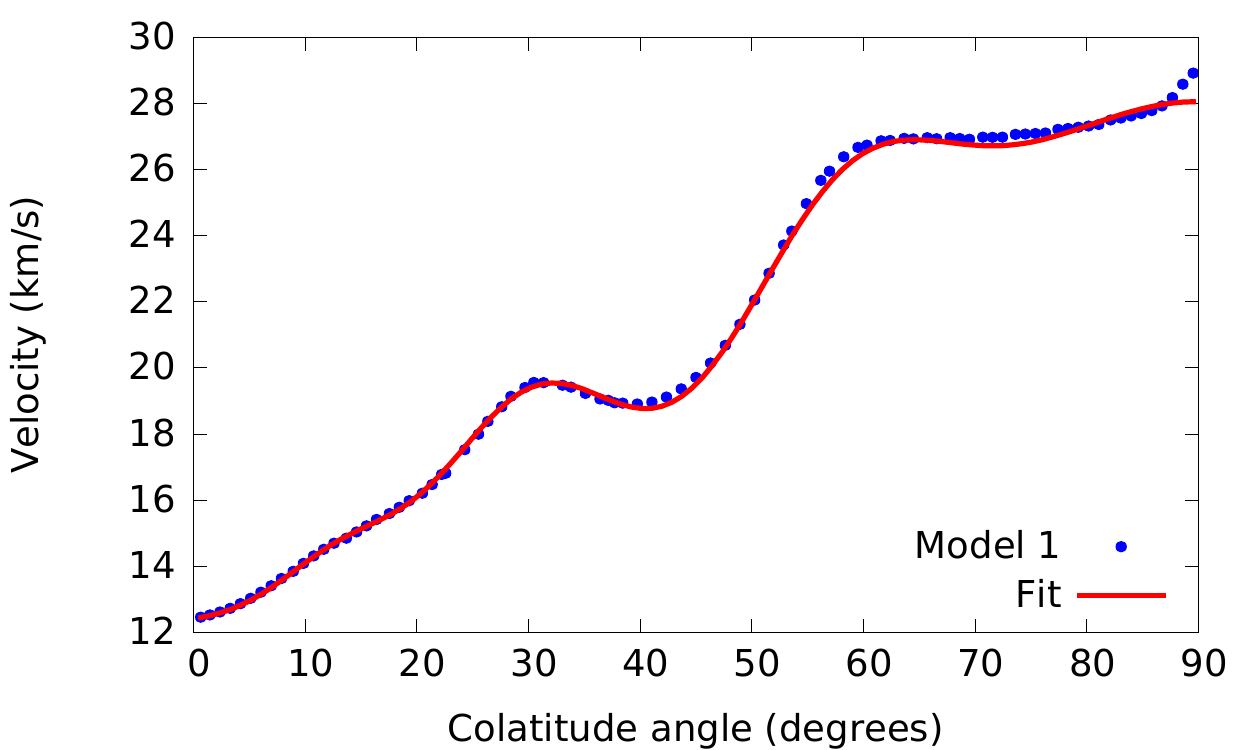} %\label{fit2}%
	\caption{Time-averaged density (top) and outflow velocity (bottom) profiles, as functions of the colatitude angle, at $20.0~$au from the barycentre of the system for our fiducial model ($\bullet$ symbols). Solid lines are analytical fits from equations~(\ref{eq9}) and~(\ref{eq10}).} %$4300\,\textrm{R}_{\sun}$
	\label{plotfit}
\end{figure}

\section{Discussion}

%GGS
%Our
One of the most important feature from the above results is that 
the outflow has a complex structure. Our results show that the time-averaged density is 
maximal at the orbital plane and minimal along the poles.  
Thus for example, the interactions of an ejected CE 
will be affected by this gas distribution, even more  the subsequent PNe formation 
\citep{2018ApJ...860...19G}. 
This suggests that the connection between PNe symmetry axis and binary star parameters 
\citep{2012MNRAS.420.2271J} is established before RLOF and survives the CE phase. 

%GGS:
The pole-to-equator density ratio is also important in the shaping of a hot bubble resulting from the shocked material by a jet from the secondary star, at the RLOF phase \citep[and references therein]{2019MNRAS.488.5615S}. For the above reasons, the 
density ratio has to be taken into account to study the correct hydrodynamical evolution of the binary system.

For a single AGB star with pulsation period $P=300~\textrm{days}$, parametrizations for mass-loss rates from \citet{1993ApJ...413..641V, 2010A&A...523A..18D} agree within one order of magnitude with the mass-loss rate of Model 3, where the binary star has its largest orbital separation and the companion star has the lowest influence. For the other models, the mass-loss rates are larger and comparable to those in AGB stars undergoing superwinds \citep{1981ASSL...88..431R, 1993ApJ...413..641V}.

There are three implications for the above result. First, the AGB timescale could be shortened if such mass-loss rates are maintained or enhanced by the CE \citep[see e.g.][]{2011MNRAS.411.2277D}. Second, superwinds at the tip of the AGB phase could be actually gravity-enhanced winds in binary systems approaching RLOF \citep[see e.g.][]{2019NatAs...3..408D}. Third, 
%GGS the primary star 
the stellar evolution of the primary 
%GGS: is 
could be driven by the presence of a companion star rather than by intrinsic properties \citep[see e.g.][]{2012Sci...337..444S}. This can reduce the final mass of white dwarfs in binary systems.

Dynamical interactions between circumbinary disks and binary stars are considered by \citet{1991ApJ...370L..35A, 2013A&A...551A..50D} as a mechanism to increase the orbit eccentricity. However our models show that there is no gravitationally-bounded circumbinary disks when radiation pressure on dust is taken into account. This implies that the above-mentioned mechanism may not work for AGB binaries with dusty winds and approaching RLOF. 
Our models also suggest that such disks observed in post-AGB binaries \citep{2006MmSAI..77..943V} must be formed after RLOF or when the dust-driven wind is over.

The accretion disk around the main-sequence companion and the Roche lobe of the latter has approximately the same radius \citep{1977ApJ...216..822P}. Using equation~(\ref{eqrl}) with the stellar mass ratio $q=M_2/M_1=0.8~M\textrm{\sun}/2.2~M\textrm{\sun}$ and $a=674.3~\textrm{R}_{\sun}$, the Roche lobe radius for the secondary star is $\sim 200~\textrm{R}_{\sun}$, which is $\sim11$ times the size of a mesh cell. At this distance the smoothed gravitational potential is identical to the Newtonian potential, and mass loss through the Lagrangian point $L_2$ is unaffected. In addition, our simulations of binary systems with the same orbital separation but more massive secondary stars (with a deeper gravitational potential well) show that the outflow structure is well described by the same formulae. Therefore we expect that our qualitative results remain valid, even though the accretion disk physics is not resolved.

Finally, our models show outflows with pole-to-equator density ratios from $10^2$ up to $\sim10^5$, however these values are reduced if radiative cooling is absent. This implies that a complete computation of PNe formation and evolution must include a self-consistent treatment of radiative cooling.

The time-averaged structure of the outflows from our interacting binary stars is well described using the proposed analytical formulae. 
Motivated by the fact that our models include both semi-detached and detached binaries, a future work will explore if our formulae remain valid for a much larger combination of orbital separations and stellar mass ratios.

\section{Conclusions}
We have 
%GGS: numerically 
computed 
numerically the outflow from interacting binary stars. The results of our study can be summarized as follows: 
mass loss from an interacting binary system hosting an AGB star is produced mainly through the outer Lagrangian points, as shown also in \citet{1989ApJ...339..268S, 1999ApJ...523..357M, 2007ASPC..372..397M}. 

The resultant pole-to-equator density ratios and mass-loss rates increase in binary systems with smaller orbital separations or with larger mass ratios $q=M_2/M_1$ (top panels of Figures~\ref{fig4} and~\ref{fig5}). At the RLOF phase, both density ratio and mass-loss rate have 
the largest values. 

These results agree with the hypothesis of binary stars as a shaping mechanisms of aspherical PNe.

If radiation pressure on dust grains is included, the gas leaving the Lagrangian $L_2$ point does not establish an outer ring of gravitationally bound matter when $q=M_2/M_1>0.78$. This behaviour is the opposite when radiation pressure is not included \citep{1979ApJ...229..223S, 2016MNRAS.455.4351P, 2017MNRAS.468.4465C}.

We find analytical formulae describing the outflow structure in terms of distance from the barycentre and the colatitude angle.
%GGS: from the unit normal vector to the orbital plane.
The formulae can be used in future studies to setup hydrodynamic simulations of CE evolution and the formation of planetary nebulae.

\section*{Acknowledgements}
We thank the anonymous referee for her/his valuable comments, which have improved considerably the article.
We thank W. J. Schuster for improving the English of our paper.   
We thank  J. A. Esquivel 
%GGS \citep{2010ApJ...725.1466E} 
and J. C. Toledo 
%GGS \citep{2014MNRAS.437..898T} 
%GGS: en los agradecimientos no se ponen referencias
for making their code available. 
This work is supported by CONACyT grant 512173, 178253 and DGAPA-PAPIIT 104017. 
\textit{Software credit}: \textsc{VisIt} \citep{HPV:VisIt}.



\bibliographystyle{mnras}
\bibliography{biblio} % if your bibtex file is called example.bib

\appendix \section{Curve fitting}

\begin{figure}
	\centering
	\includegraphics[width=1.0\columnwidth]{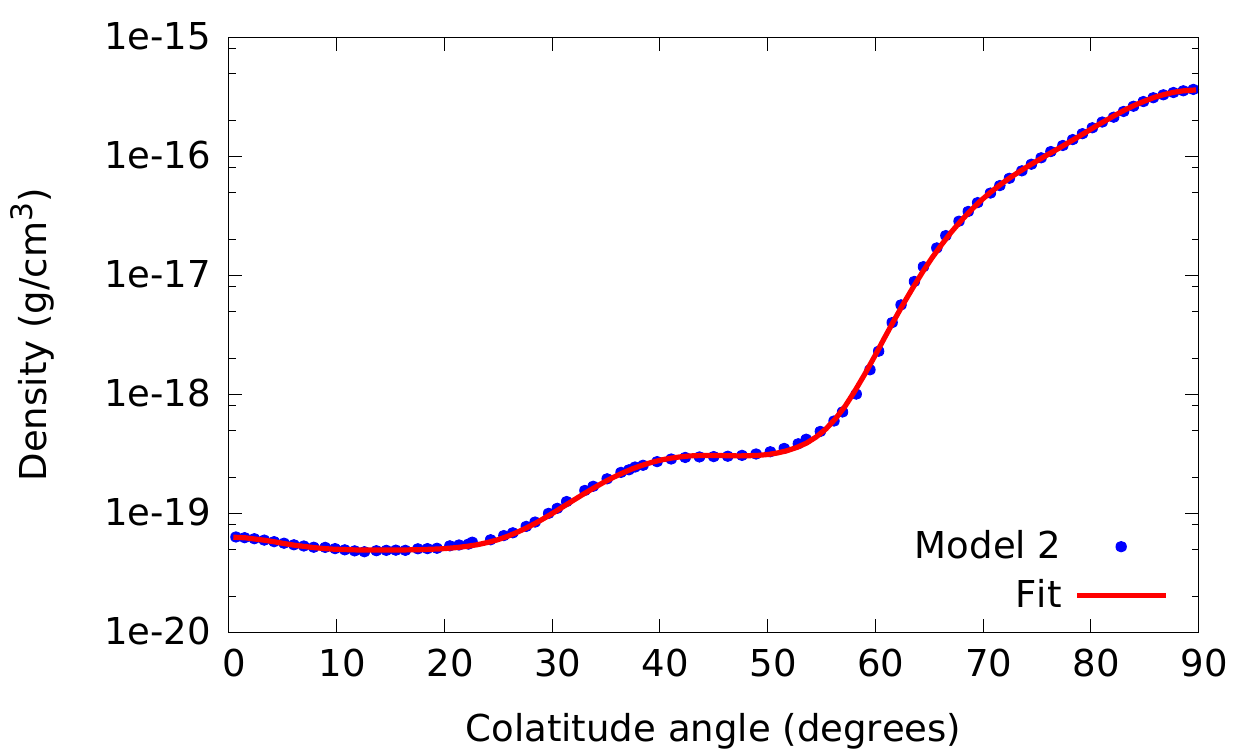} %\label{fit}
	\hfill
	\includegraphics[width=1.0\columnwidth]{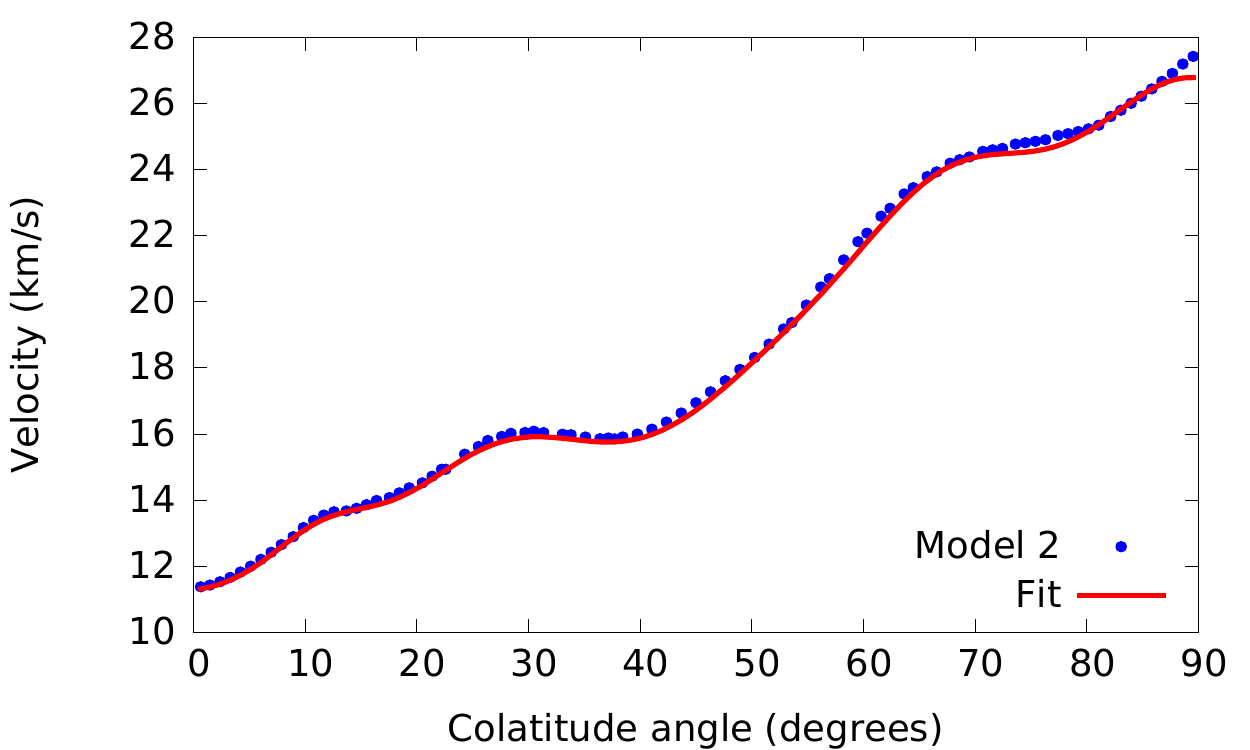} %\label{fit2}%
	\caption{Same as Fig.~\ref{plotfit} for model 2.}
	\label{plotfit2}
\end{figure}

\begin{figure}
	\centering
	\includegraphics[width=1.0\columnwidth]{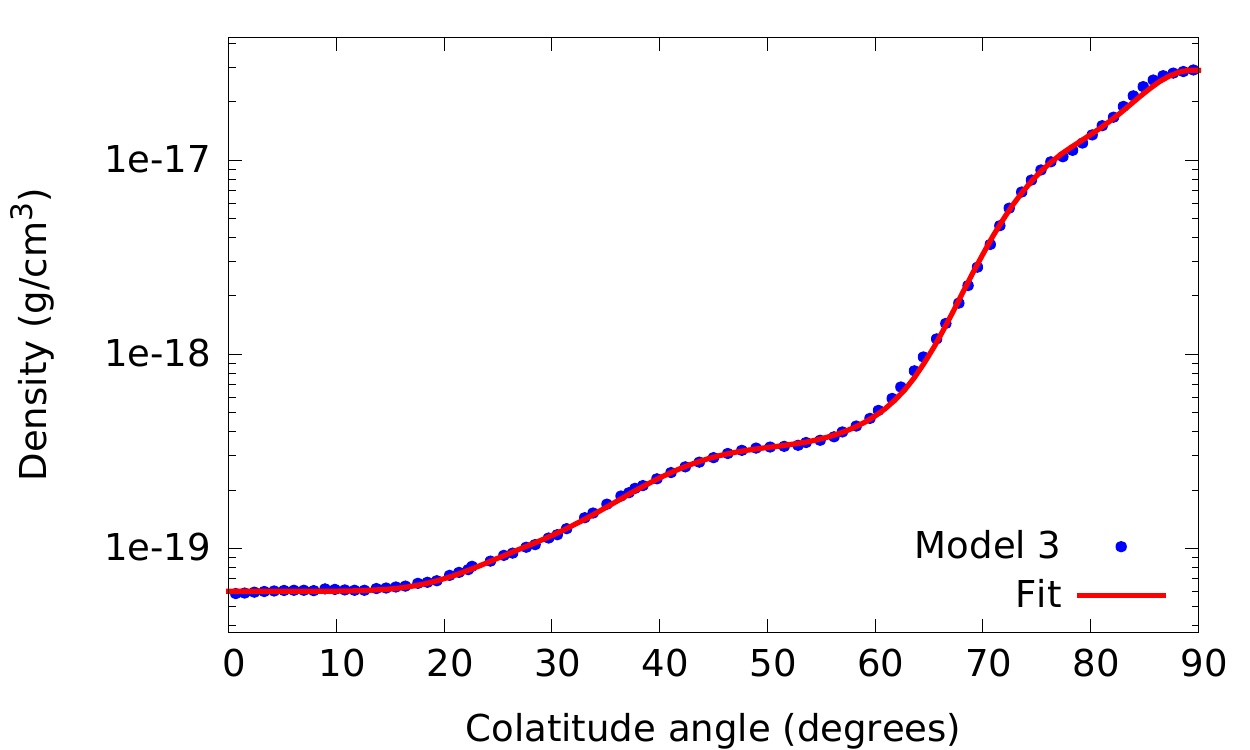} %\label{fit}
	\hfill
	\includegraphics[width=1.0\columnwidth]{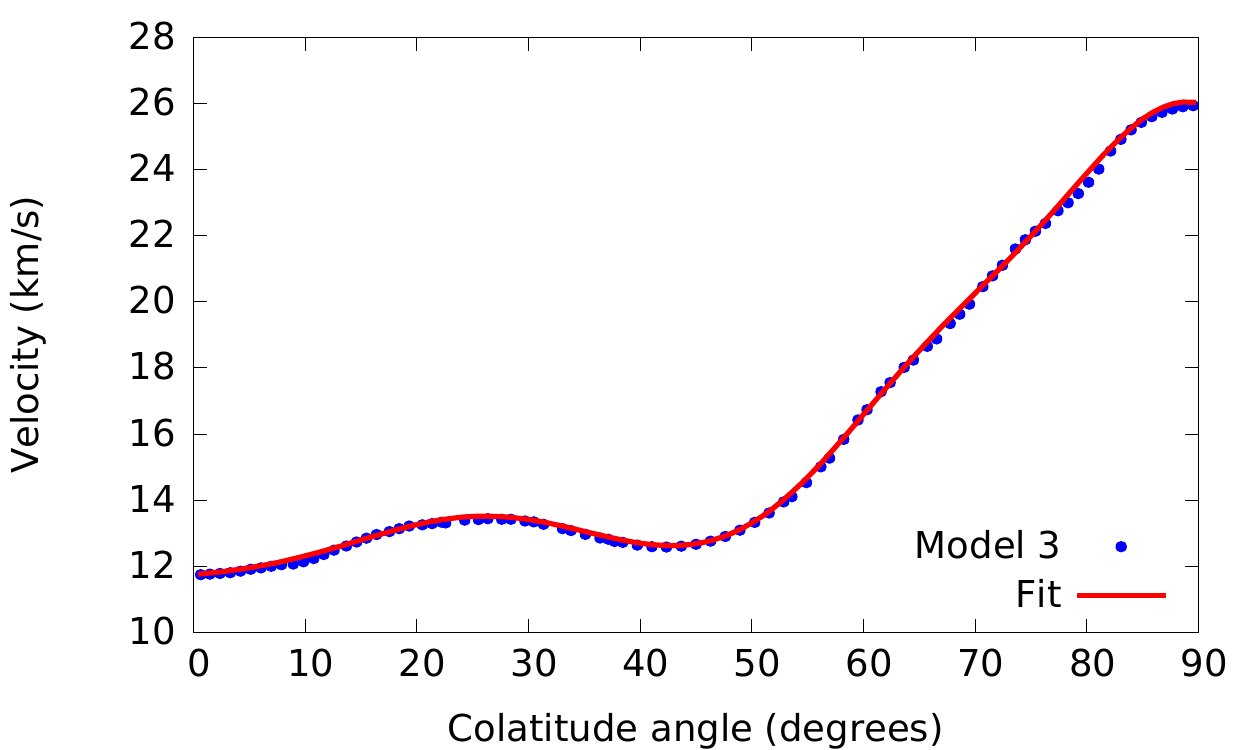} %\label{fit2}%
	\caption{Same as Fig.~\ref{plotfit} for model 3.}
	\label{plotfit3}
\end{figure}

\begin{figure}
	\centering
	\includegraphics[width=1.0\columnwidth]{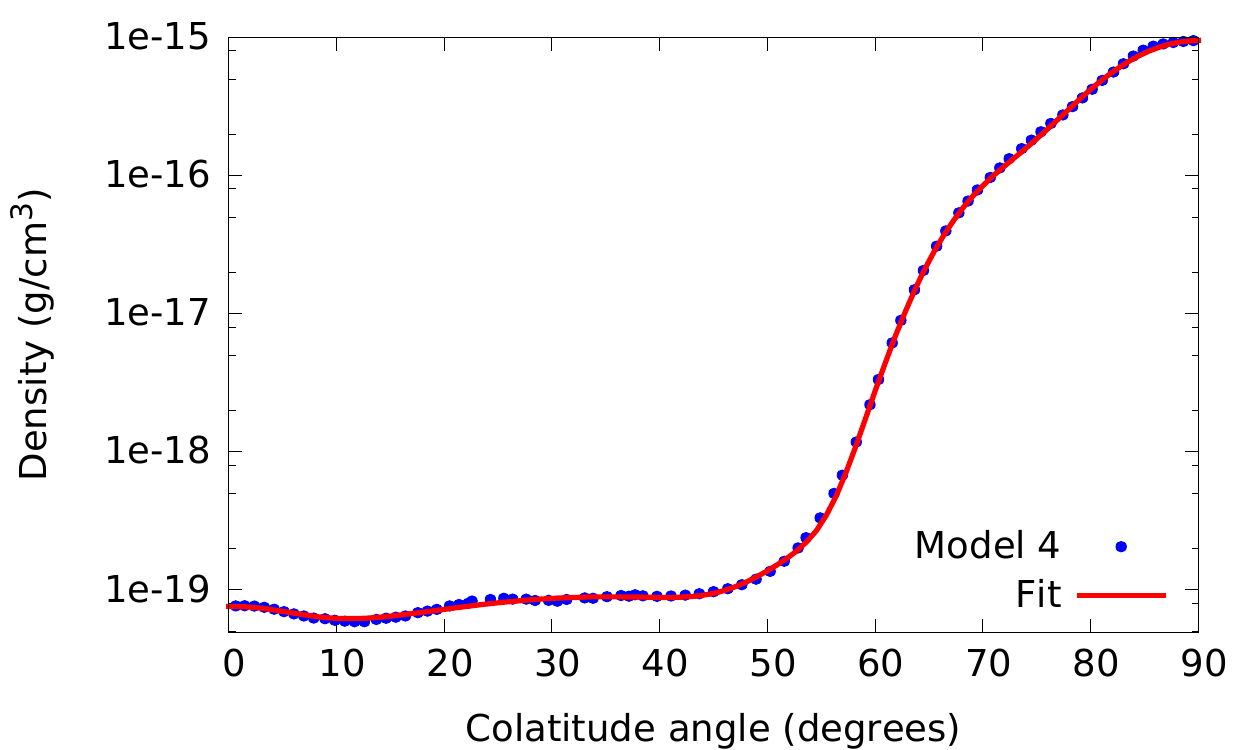} %\label{fit}
	\hfill
	\includegraphics[width=1.0\columnwidth]{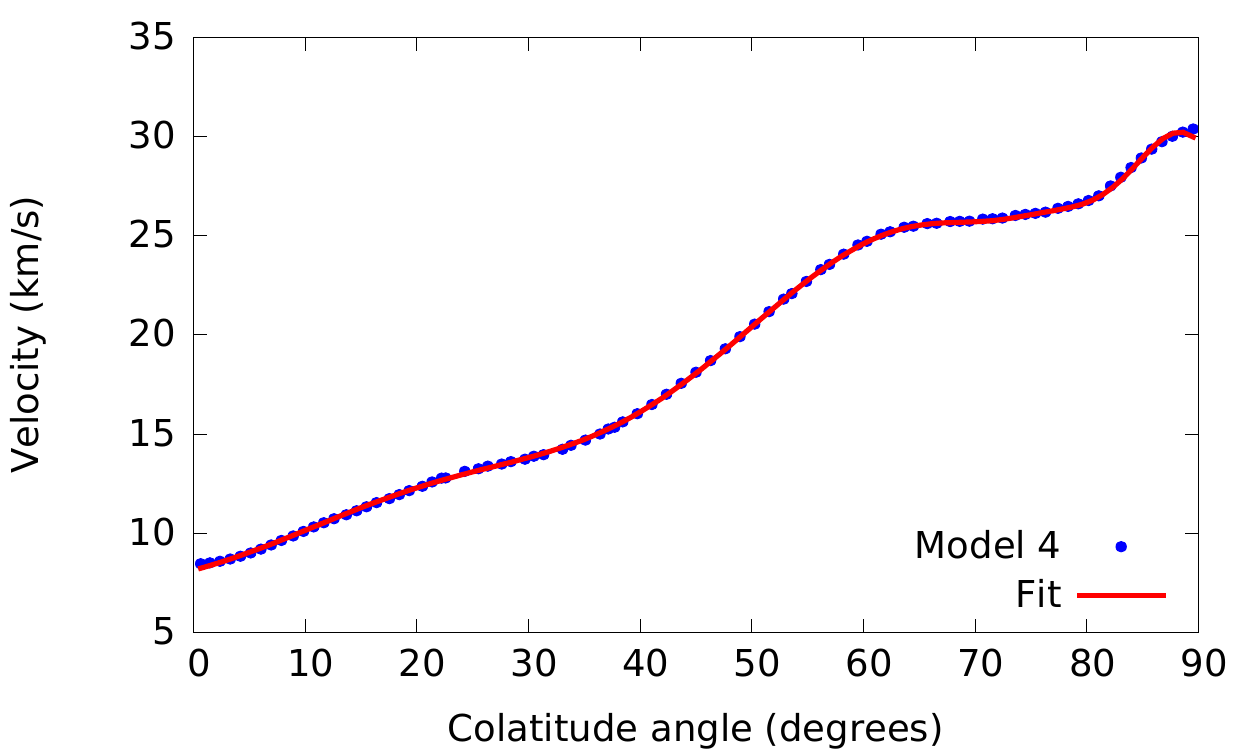} %\label{fit2}%
	\caption{Same as Fig.~\ref{plotfit} for model 4.}
	\label{plotfit4}
\end{figure}

\begin{figure}
	\centering
	\includegraphics[width=1.0\columnwidth]{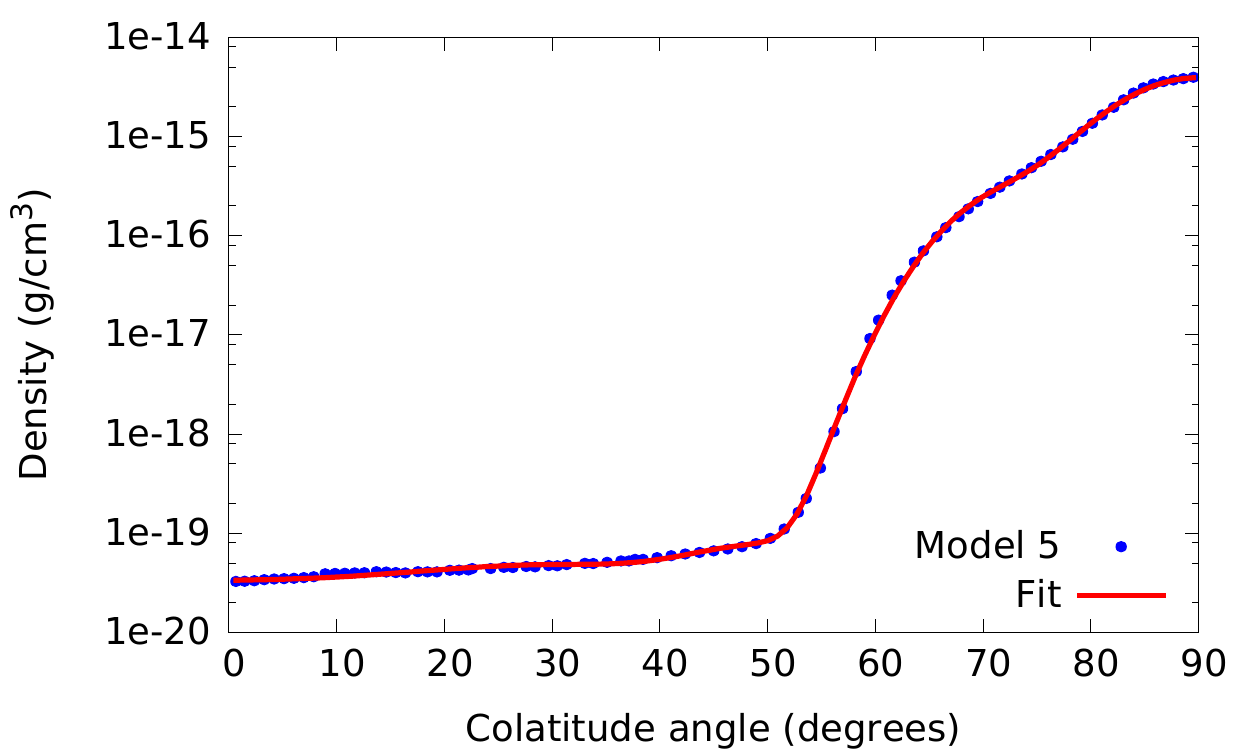} %\label{fit}
	\hfill
	\includegraphics[width=1.0\columnwidth]{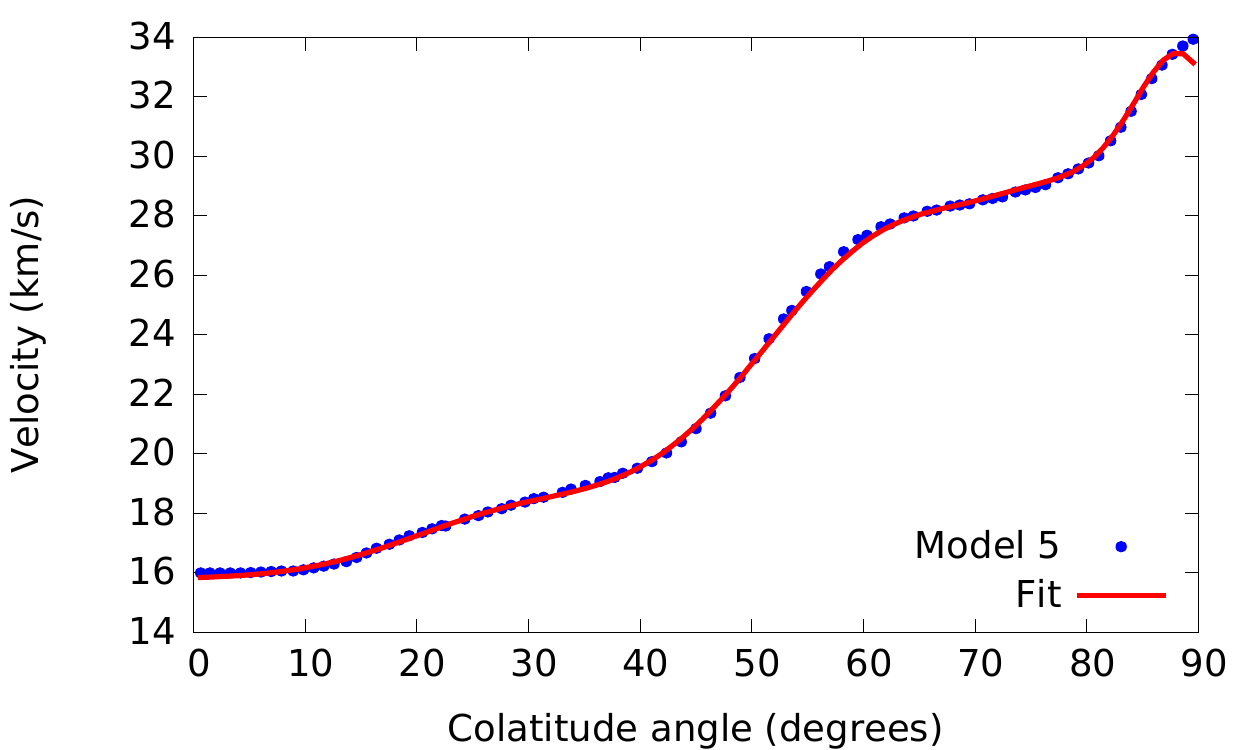} %\label{fit2}%
	\caption{Same as Fig.~\ref{plotfit} for model 5.}
	\label{plotfit5}
\end{figure}

\begin{figure}
	\centering
	\includegraphics[width=1.0\columnwidth]{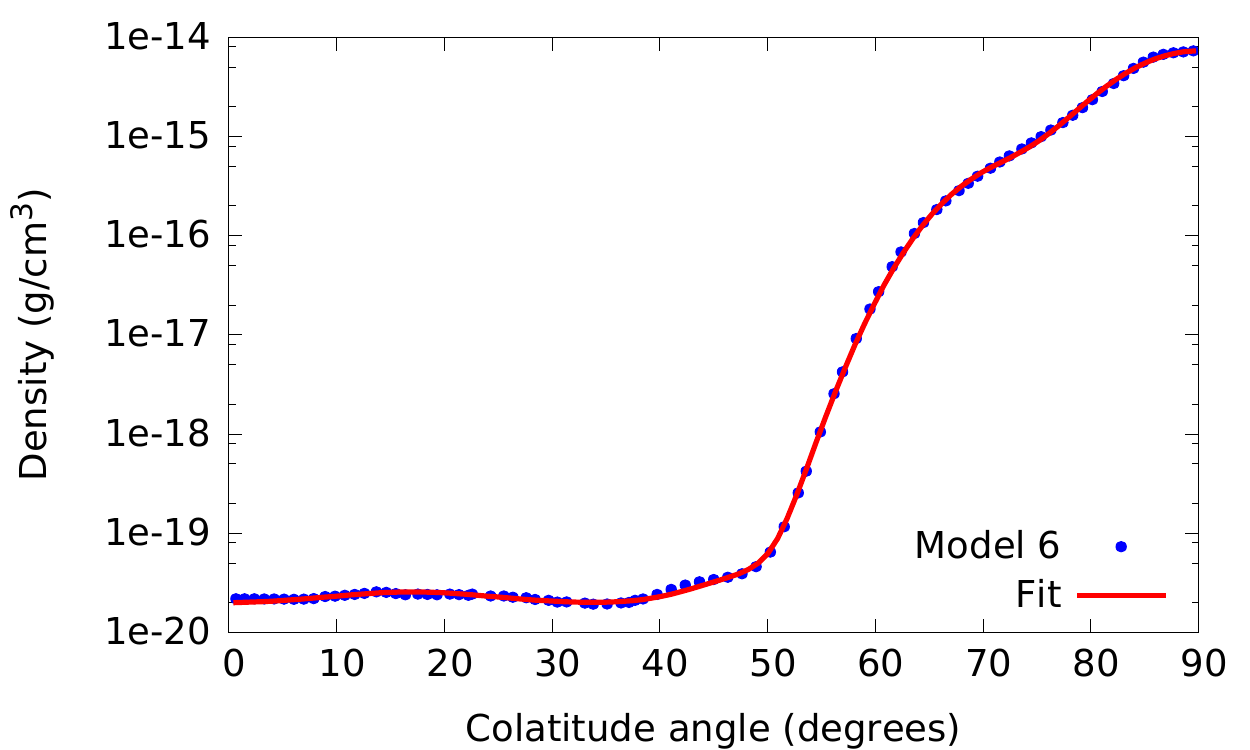} %\label{fit}
	\hfill
	\includegraphics[width=1.0\columnwidth]{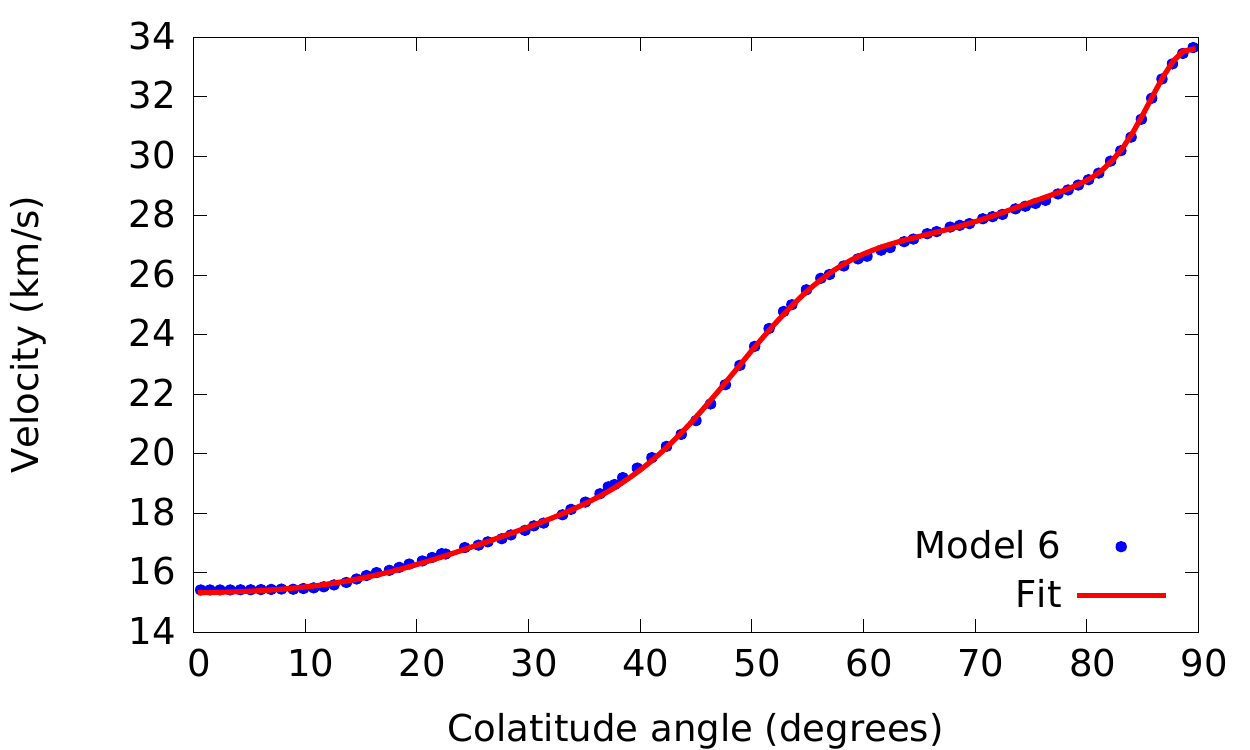} %\label{fit2}%
	\caption{Same as Fig.~\ref{plotfit} for model 6.}
	\label{plotfit6}
\end{figure}

\begin{figure}
	\centering
	\includegraphics[width=1.0\columnwidth]{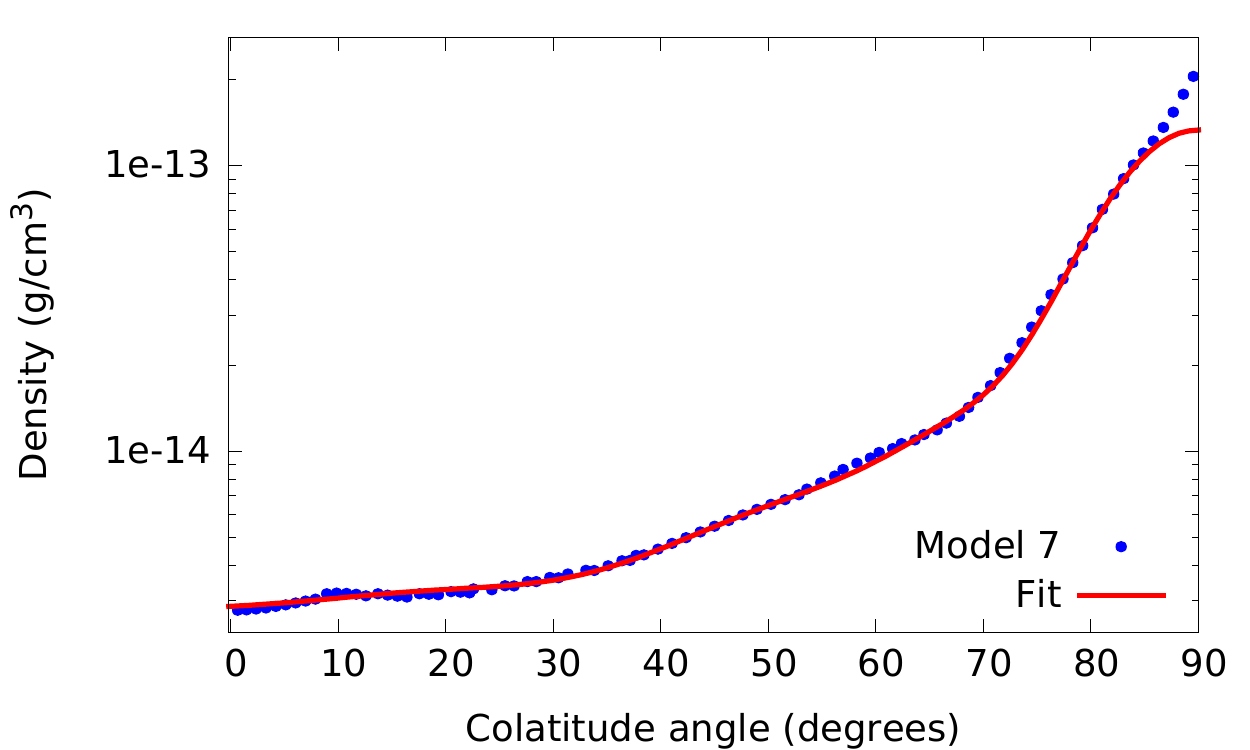} %\label{fit}
	\hfill
	\includegraphics[width=1.0\columnwidth]{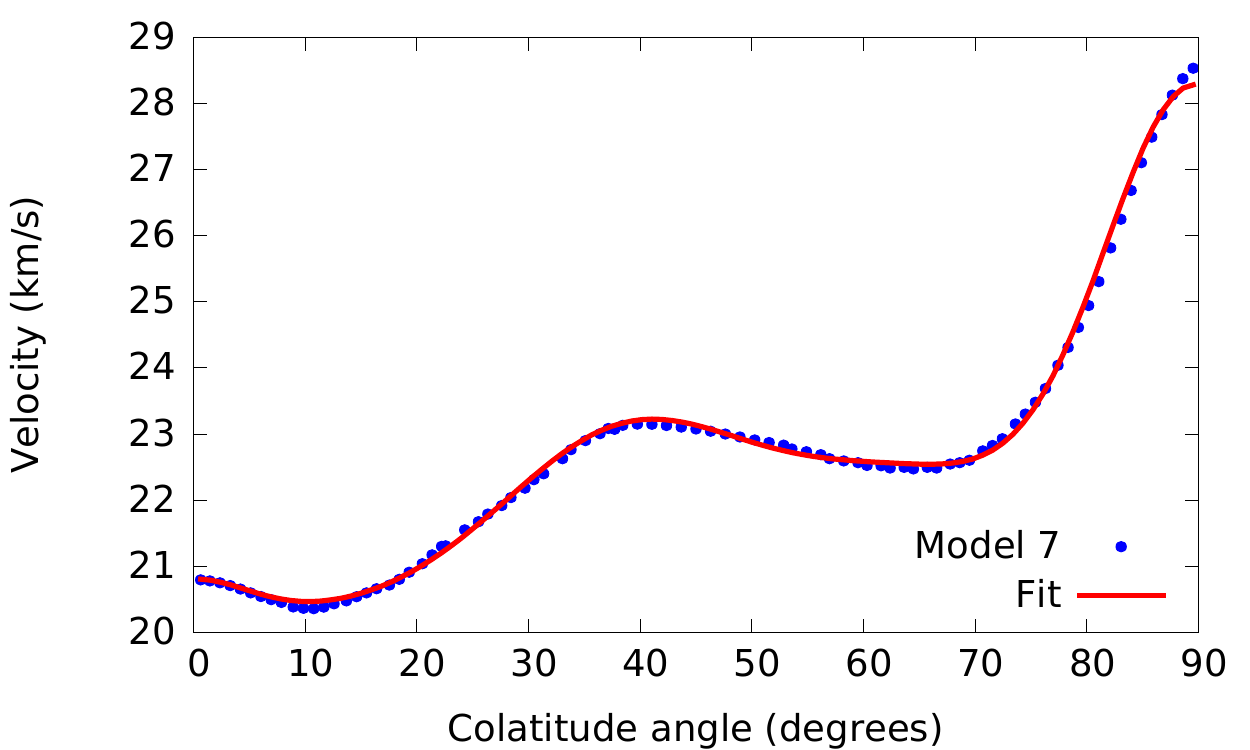} %\label{fit2}%
	\caption{Same as Fig.~\ref{plotfit} for model 7.}
	\label{plotfit7}
\end{figure}

% Don't change these lines
\bsp	% typesetting comment
\label{lastpage}
\end{document}